\documentclass{article}
\usepackage{PRIMEarxiv}
\usepackage[utf8]{inputenc} 
\usepackage[T1]{fontenc}    
\usepackage{yfonts}
\usepackage{hyperref}       
\usepackage{url}            
\usepackage{booktabs}       
\usepackage{amsfonts}       
\usepackage{nicefrac}       
\usepackage{microtype}      
\usepackage{lipsum}
\usepackage{fancyhdr}       
\usepackage{graphicx}       
\graphicspath{{figures/}}     

\newcommand\BibTeX{{\rmfamily B\kern-.05em \textsc{i\kern-.025em b}\kern-.08em
		T\kern-.1667em\lower.7ex\hbox{E}\kern-.125emX}}

\usepackage{moreverb}
\usepackage{bm}
\usepackage{amsmath}
\usepackage{cleveref}
\usepackage{bbm}
\usepackage{threeparttable,booktabs}

\newcommand{\Normal} {{\mathcal N}}

\newcommand{\param} {\Theta}
\newcommand{\Data}  {\mathcal{D}}
\newcommand{\Ypred} {\boldsymbol{Y}}
\newcommand{\paramDis} {\mathbf{m}}
\newcommand{\meanex} {\overline{\gamma_r}}

\newcommand{\etal}{\textit{et al}. }

\title{Detecting Outbreaks Using a Latent Field: Part II - Scalable Estimation
}

\author{
	Wyatt Bridgman \\
	Sandia National Laboratories \\
	Livermore, CA \\
	\texttt{whbridg@sandia.gov} \\
	\And
	Cosmin Safta \\
	Sandia National Laboratories \\
	Livermore, CA\\
	\texttt{csafta@sandia.gov} \\
	\And
	Jaideep Ray \\
	Sandia National Laboratories \\
	Livermore, CA \\
	\texttt{jairay@sandia.gov} \\
}

\begin{document}
	\maketitle

\begin{abstract}
In this paper, we explore whether the infection-rate of a disease can serve as a robust monitoring variable  in epidemiological surveillance algorithms. The infection-rate is dependent on population mixing patterns that  do not vary erratically day-to-day; in contrast, daily case-counts used in contemporary surveillance algorithms are corrupted by reporting errors. The technical challenge lies in estimating the latent infection-rate from  case-counts. Here we devise a Bayesian method to estimate the infection-rate across multiple adjoining areal  units, and then use it, via an anomaly detector, to discern a change in epidemiological dynamics. We extend  an existing model for estimating the infection-rate in an areal unit by incorporating a Markov random field model, so that we may estimate infection-rates across multiple areal units, while preserving spatial correlations observed in the epidemiological dynamics. To carry out the high-dimensional Bayesian inverse problem, we develop an implementation of mean-field variational inference specific to the infection model and integrate it with the random field model to incorporate correlations across counties. The method is tested on estimating the COVID-19 infection-rates across all 33 counties in New Mexico  using data from the summer of 2020, and then employing them to detect the arrival of the Fall 2020 COVID-19 wave. We perform the detection using a temporal algorithm that is applied county-by-county. We also show how the infection-rate field can be used to cluster counties with similar epidemiological dynamics.
\end{abstract}

\keywords{VI, variational inference; ELBO, evidence lower bound; COVID-19; New Mexico; outbreak detection}

\maketitle



%
%
\section{Introduction}
\label{sec:intro}
There have been many attempts to estimate the infection-rate of an outbreak~\cite{DazaTorres:2022,WangZ:2020,ChenP:2021}, especially for the COVID-19 pandemic, including our own work~\cite{Blonigan:2021,Lin:2021,Safta:2021}. In these studies, the infection-rate is modeled as a time-varying function, which is estimated by fitting a disease model to observed data e.g., case-counts of symptomatic (diagnosed) individuals per day. The infection-rate is used to forecast the outbreak over a small time horizon, e.g., two weeks, and then compared with the data from that period. This ability to compare observed case-counts with model forecasts can be used to fashion an outbreak detector for disease surveillance - if the model forecasts and reported case-counts do not match, it could indicate a change in the epidemiological dynamics, either due to a change in human mixing patters e.g., due to lock-downs~\cite{Safta:2021} or due to a new variant of the pathogen~\cite{Blonigan:2021}. A shortcoming of these studies is that they do not contain any spatial information on the variation of the infection-rate, which impairs their usefulness in disease surveillance - public health policy is determined globally (e.g. for a nation) and then adapted and applied locally (e.g., in a county). Being able to estimate an infection-rate \emph{field}, defined over multiple areal units (e.g., counties),  can thus be very helpful. However, the task is challenging, as the problem now requires one to use epidemiological data e.g., case-counts, collected from each areal unit, which could have a small population. Such data, gathered from small populations, tends to be contaminated with high-variance noise (reporting errors) and has to be compensated for by imposing the spatial patterns extant in epidemiological dynamics caused by mixing of humans between neighboring areal units. In the Part I of this paper~\cite{24sr3a,safta2024detecting} (henceforth \emph{Part I paper}), we formulated and solved an inverse problem for estimating the infection-rate field and demonstrated it on COVID-19 case-count data collected between June 1, 2020 and September 15, 2020 from the counties of New Mexico, USA. We used the infection-rate to devise an outbreak detector to detect the start of the Fall 2020 wave of COVID-19 infections in New Mexico (NM), and determined that it did so about a week earlier than a conventional outbreak detector. The method employs a separate infection-rate parameterization inside each areal unit, but links them together via a Gaussian Markov random field (GMRF) to impose spatial correlations. The parameters are estimated using an adaptive Markov Chain Monte Carlo (AMCMC~\cite{Haario:2001}) sampler. Since AMCMC is not very scalable in terms of the number of parameters that can be estimated, we were limited to a study over three adjoining NM counties - Bernalillo, Santa Fe and Valencia.

In this paper, we adopt the formulation of the inverse problem from our previous paper~\cite{24sr3a,safta2024detecting} and adapt it for high-dimensional inversion. Our approach is based on the hypothesis that scalable mean-field variational inference (MFVI~\cite{Han:2019}) will allow us to obtain a useful estimate of the infection-rate field, defined over many areal units, despite the approximations inherent in MFVI. In MFVI, one imputes a parameterized posterior distribution for the infection-rate field; in our case, we model the posterior distribution of (transformations of) our parameters as independent Gaussians of unknown means and standard deviations. These means and standard deviations are estimated by minimizing an objective function using an iterative, gradient-based algorithm. We will test our method by reconstructing the time-dependent infection-rate field over all 33 counties of NM, with case-count data of diverse qualities, collected between June 1 and September 15, 2020. We will also fashion an outbreak detector using the infection-rate fields and check its performance in detecting the Fall 2020 COVID-19 wave against the one devised in our Part I paper~\cite{24sr3a,safta2024detecting} that uses AMCMC to perform the estimation. In doing so, we will address the following research questions:
\begin{itemize}
    \item How does the approximate posterior distribution employed in MFVI compare against the ``true'' one (computed using AMCMC) from our Part I paper? By necessity, this comparison will be limited to the three counties included on our previous paper.
    \item The spatial correlations embedded in the field estimation problem allow an arial unit to ``borrow information'' from its neighbors and compensate for poor-quality data. How robust is this correlation i.e., when it fails due to poor quality data, does it do so catastrophically (i.e., the estimation process stops) or gracefully, with a non-informative (or erroneous) estimate?
    \item How does the approximate infection-rate field estimated by MFVI affect the accuracy of the outbreak detector? Is MFVI sufficiently accurate to be useful, despite its inherent approximations? Further, does the infection-rate field, estimated over all NM counties, reveal anomalous spatial structures during the start of the Fall 2020 wave?
\end{itemize}

The paper has three main contributions. Our first contribution is the determination of the degree to which the predictive skill of the disease model forecast is affected by the use of an imputed and approximate posterior distribution; we find that it is not much vis-\`{a}-vis the AMCMC solution. Our second contribution is the discovery that despite the approximations inherent in MFVI, the infection-rate field so estimated retains sufficient information on the outbreak to detect epidemiological anomalies. Note that we will not attempt to make a proper outbreak detector in this paper; that is left to future work. Our third contribution is of a numerical nature. Our disease model includes a convolutional integral making it difficult (and expensive) to compute the gradient of the objective function, analytically or by finite differences. Our innovations lie in how the convolution and the gradient are computed numerically via quadrature and the reparametrization of the objective function that allows us to use an unconstrained optimization method, despite the need to preserve non-negativity of parameters being estimated.

The paper is structured as follows. In \S~\ref{sec:litrev}, we review relevant literature. In \S~\ref{sec:form} we formulate the inverse problem and the MFVI adaptation, and results are presented in \S~\ref{sec:res}. Outbreak detection is explored \S~\ref{sec:discussion}. We conclude in \S~\ref{sec:concl}.

\section{Literature Review}
\label{sec:litrev}

In this section, we review some of the literature that undergird our spatiotemporal epidemiological model, as well as work on variation inference algorithms that will be used to scale our model to high-dimensional inverse problems involved in estimating infection-rate fields.

{\bf Spatial modeling in epidemiology: } Epidemiological dynamics show spatial autocorrelation because of human mixing as well as the dependence of some outbreaks on socioeconomic and demographic covariates which do not change erratically in space or time. The COVID-19 pandemic was observed and recorded with fine spatiotemporal granularity, and the data has been subjected to much spatiotemporal modeling.  Many such studies found that the spread of the disease was mediated mostly by human mixing, rather than by socioeconomic factors; Huang~\etal~\cite{Huang:2021} found it to be so in Hubei province in China, and we found much the same result for New Mexico in our Part I paper~\cite{24sr3a,safta2024detecting}. Geng~\etal~\cite{Geng:2021} analysed US data and found that spatial patterns' lengthscales ranged from the county-level to the nation; similar results were found by Schuler~\etal~\cite{Schuler:2021} for Germany. McMahon~\etal~\cite{McMahon:2022} analyzed data at the county-level and found that the correlation lengthscales for spatial variability changed during the course of the pandemic; further, the correlation in epidemiological dynamics between urban centers were stronger than elsewhere. Indika~\etal~\cite{Sathish:2023} analyzed data from the counties of Virginia and found that spatial autocorrelation of case-counts, as quantified by Moran test statistics, were impacted by, and linked to, executive orders at the state level. Thus, COVID-19 has presented us with much evidence of spatial auto-correlation in epidemiological dynamics.

The incorporation of spatial autocorrelation in the modeling and estimation of infection-rate fields is rare; however, it has been extensively used in disease mapping. In disease maps, one develops a field, called \emph{relative risk} $r_k$, that is used to adjust an expected value of morbidity $e_k$ to the locally observed  case-counts $y^{(obs)}_k$ in an areal unit (e.g., county) $k$, often via a Poisson or Negative Binomial link i.e., $y^{(obs)}_k \sim {\mathrm{ Poisson}}(r_k e_k)$. The expected value $e_k$ is usually obtained from a regional average. The relative risk $r_k$ is modeled using covariates of disease activity e.g., socioeconomic conditions $\log(r_k) = \boldsymbol{z}_k \cdot \boldsymbol{\beta} + \phi_k$, where $\boldsymbol{z}_k$ are co-variate risk factors for areal unit $k$, $\boldsymbol{\beta}$ are regression weights and $\phi_k$ captures auto-correlated random effects in space using a random field model. The simplest random field model is iCAR (intrinsic Conditional AutoRegressive~\cite{Lawson:2017}), a specific type of Gaussian Markov Random Field (GMRF). Thus 
\[
\phi = \{\phi_k\} \sim \Normal\left(0, \{\tau^2 Q\}^{-1} \right), \mbox{\hspace{1cm}} Q = {\rm diag}(W\boldsymbol{1}) - W,
\]
where $W$ is the adjacency matrix of the areal units (i.e., $w_{ij} = 1$ if areal units $i$ and $j$ share a boundary). The object of estimation from data is $\tau^2$. The precision matrix $Q$ tends to be sparse. Another common model is the Besag-York-Mollie (BYM) model~\cite{Besag:1991} which decomposes $\phi_k$ as $\phi = \phi^1 + \phi^2, \phi^1 \sim \Normal(0, \{\tau^2 Q\}^{-1})$ and $\phi^2 \sim \Normal(0, \sigma^2I)$. An adaptation of the BYM was used in the Part I paper and is also used in this paper. More details on spatial models is available in our Part I paper, where we develop the spatial model employed in the current paper.

The work done by Lawson and collaborators~\cite{10ls2a,23la1a,23kl6a} is the closest to our own. Fundamentally we extend an older model of ours~\cite{Safta:2021,Blonigan:2021}, meant for a single areal unit, to encompass multiple areal units. Each areal unit contains its own parameterized infection-rate representation, but are ``stitched'' together using a BYM model. The parameters are estimated by solving an inverse problem, conditioned on case-count data from areal units. Our model also includes a model for the incubation period. In contrast, Lawson and co-workers model case-counts directly; the clearest exposition of the model in in Lawson and Song, 2010~\cite{10ls2a}, and it has been used with COVID-19 data from South Carolina~\cite{21lk2a} and the UK~\cite{21sl3a}. Lawson and co-workers, much like us, have used the disagreement between calibrated model forecasts and data as signs of epidemiological anomalies and devised metrics such as the Surveillance Kullback-Liebler~\cite{16rl2a} (SKL) and Surveillance Conditional Predictive ordinate~\cite{11cl2a} (SCPO) to detect them.

{\bf Part I paper: } Our current paper is an algorithmic follow-on to our Part I paper~\cite{24sr3a,safta2024detecting}, which focused on developing a spatial model, with a latent (parameterized) infection-rate field, where the parameters were estimated from COVID-19 data from the counties of NM. As mentioned above, it consists of a parameterized infection-rate model for each areal unit~\cite{Safta:2021,Blonigan:2021} (i.e., a NM county) which are linked to each other via a (modification of) BYM model, much like the work by Lawson and collaborators~\cite{10ls2a,23la1a,23kl6a}. The spatial dependence is between nearest-neighbors only; the details of how this was arrived at are in the Part I paper. The model was fitted to data from three adjoining NM counties - Bernalillo, Santa Fe and Vaelncia - using AMCMC; its lack of scalability prevented us from broadening the model to more counties. It therefore  motivated the current paper, where we use scalable, but approximate, variational inference to address the problem of infection-rate estimation and forecasting across all 33 NM counties. As in Lawson's work~\cite{16rl2a,11cl2a} we developed metrics to detect anomalous epidemiological behavior, cast as a disagreement between forecasts (using a calibrated model) and data. Specifically, we used the forecating capability to detect the arrival of the Fall 2020 wave using data collected up to September 15, 2020, and had no diffculties in the detection within a week of its arrival. We also tested the anomaly detector on data collected up to August 15, 2020, long before the arrival of the COVID-19 wave, to check the detector's susceptibility to false positives.

{\bf Variational Inference: } Inverse problems for model calibration often require the approximation of intractable probability densities arising from Bayesian inference. A standard approach based on sampling is Markov Chain Monte Carlo (MCMC)~\cite{Brooks:1998} but suffers from scalability issues due to slow convergence rates for high-dimensional and/or multimodal distributions~\cite{Ravenzwaaij:2018,Roberts:1996}. Variational inference (VI)~\cite{Blei:2017} provides an alternative to sampling techniques where approximate inference is recast as seeking a member of a family of approximating densities which minimizes a discrepancy measure such as KL-divergence. Originally developed for probabilistic graphical models \cite{Jordan:1999} where some degree of analytical tractability is maintained, it has more recently been extended to many-parameter models, such as those seen in deep learning, through gradient-based iterative schemes adapted to a probabilistic setting~\cite{Blundell:2015,Kingma:2019,Hernandez:2015}. These techniques are often termed Stochastic Variational Inference (SVI) and can exploit the automatic differentiation available in large ML models. They offer significantly improved scalability over MCMC while potentially sacrificing some approximation quality depending on the set of approximating distributions used. VI has seen successful applications to a number high-dimensional inverse problems in areas including medical classification \cite{Tanno:2017,Rkaczkowski:2019} and segmentation \cite{Ozdemir:2017,Luo:2020}, computer vision and image processing \cite{Carvalho:2020}, natural language processing \cite{Liang:2007, Hu:2021}, and physics-based models \cite{Meng:2021,Yang:2021}.

In addition to this broad range of application spaces, VI has more recently been adopted in a growing number of epidemiological modeling challenges. Neural ODEs have been extended to a Bayesian setting where they can be calibrated with VI and applied to state space epidemiological models \cite{Dandekar:2020}. Model selection and dynamic causal modeling based on evolving real-world time series using VI have been applied to COVID-19 outbreaks \cite{Friston:2020, Friston:2022} to provide online forecasting tools. Dynamical system inference for spatio-temporal modeling of infectious diseases using ODE and PDE formulations has been carried out at the state scale and combined with Bayesian neural networks \cite{Wang:2020}. Model calibration with VI has also been explored using alternatives to standard state-space models such as graph-coupled Hidden Markov Models \cite{Fan:2016}. In additional to predictive model calibration, generative probabilistic modeling using VI-based approximations of distributions has also been explored to generate mission information about disease spread \cite{Biazzo:2022}.

\section{Formulation}
\label{sec:form}
Here we propose an epidemiological model to forecast infection-rates across adjacent geographical regions and use these forecasts to detect emergent outbreaks. The model is an extension of previous work by Safta~\etal\cite{Safta:2021} and Blonigan~\etal\cite{Blonigan:2021} for epidemic forecasts over a single-region wave to multi-region outbreak detection. In this section we will first describe the single-region model and then present statistical approaches to estimate the model parameters over adjacent geographical regions.

\subsection{Epidemiological Model}

The epidemiological model is defined by a spatio-temporally varying infection-rate model and an incubation model given by
\begin{eqnarray}
f_{inf}(t;t_0^r,k^r,\theta^r) &=& \frac{(\theta^r)^{-k}(t-t_0^r)^{k-1}}{\Gamma(k^r)}\exp\left(-\frac{t-t_0^r}{\theta^r}\right) \\
F_{inc}(t;\mu,\sigma) &=& \frac{1}{2}\mathrm{erfc}\left(-\frac{\log t-\mu}{\sigma\sqrt{2}}\right)
\end{eqnarray}
where the infection rate $f_{inf}$ is a Gamma distribution with shape and scale parameters $k^r$ and $\theta^r$, respectively. $F_{inc}(t;\mu,\sigma)$ is the incubation period for COVID-19, taken from Lauer \etal\cite{Lauer:2020}; further details are in the Part I paper. The parameter $t_0^r$ represents the start of the outbreak and will be inferred along with the infection-rate parameters. Note that $1 \leq r \leq R$ indexes the spatial region. The number of people that turn symptomatic over the time interval $[t_{i-1},t_i]$ is given by
\begin{eqnarray}
    y_r(i;t_0^r, N^r, k^r,\theta^r)  &=&N_r\int_{t_0^r}^{t_i} f_{inf}\left(\tau-t_0^r;k^r,\theta^r\right)
    \left[F_{inc}(t_i-\tau;\mu,\sigma) - F_{inc}(t_{i-1}-\tau;\mu,\sigma)\right]d\,\tau
    \label{eq:model-pred}
\end{eqnarray}
so that $\mathbf{y}(i) = \mathbf{y}_i = \left[y_1(i) \cdots y_{R}(i) \right]^T $ and $i$ represents the time-dependence of the predictions. Here, $N^r$ is the fourth and final region-dependent parameter and represents the total number of people infected during the entire epidemic wave in spatial region $r$ normalized by the population of region $r$.

The noisy model predictions are defined as 
\begin{equation}
    \mathbf{y}^{(o)}_i = \mathbf{y}^{(p)}_i + \boldsymbol{\epsilon}_i = \mathcal{M}(t_i; \paramDis) + \boldsymbol{\epsilon}_i, \hspace{3mm} \boldsymbol{\epsilon}_i \sim \mathcal{N}(\mathbf{0}, \mathbf{\Sigma}_i).
    \label{eq:obs_eqn}
\end{equation}
Here $\paramDis = \mathrm{vec} \left( \paramDis_r \right)$, where $\mathbf{m}_r^T = (t_0^r, N^r, k^r, \theta^r)$,  are the region specific model parameters for $r = 1,\ldots,R$ and $\mathcal{M}(:)$ is an epidemiological / disease model that uses $f_{inf}(:)$. In addition, $\mathbf{y}^{(o)}_i$ is the vector of observed case-counts on day $i$ for all $R$ regions, $\mathbf{y}^{(p)}_i$ is the vector of $R$ case-count predictions for the same day using the model $\mathcal{M}$ and $\boldsymbol{\epsilon}_i$ is the ``noise'' i.e. observed case-counts on day $i$ that cannot be explained by $\mathcal{M}$ (mostly reporting errors). To account for spatial correlations and heteroscedastic noise seen in case-counts, the noise is assumed to be composed of two terms $\boldsymbol{\epsilon}_i = \boldsymbol{\epsilon}_{i,1} + \boldsymbol{\epsilon}_{i,2}$ where the first is given by a Gaussian Markov Random Field (GMRF) model while the second represents temporally-varying, independent Gaussian noise. Letting $\Data $ and $\param$ represent the data and parameters, respectively, the likelihood then takes the form
\begin{equation}
    p(\Data  \ \vert \  \param)=\prod_{i=1}^{N_d}\frac{1}{(2\pi)^{N_r/2}
    \mathrm{det}\mathbf{\Sigma}_i^{1/2}}\exp\left(-\frac{1}{2}(\mathbf{y}^{(o)}_i-\mathbf{y}^{(p)}_i)
    \mathbf{\Sigma}_i^{-1}(\mathbf{y}^{(o)}_i-\mathbf{y}^{(p)}_i)^T\right)
    \label{eq:likelihood}
\end{equation}
where $\mathbf{\Sigma}_i$ is given by
\begin{equation}\label{eq:noise-model}
    \mathbf{\Sigma}_i = \tau_{\Phi}
    P^{-1} + \mathrm{diag}\left(\sigma_a+\sigma_m \mathbf{y}^{(p)}_i\right)^2.
\end{equation}
The first term $P = \left[D-\lambda_{\Phi}\mathbf{W}\right]$ in Eq.~\eqref{eq:noise-model} forms the precision matrix of a GMRF component of the noise where the strength of correlations induced by adjacent regions is governed by $\lambda_{\Phi}$. The relative topology of regions is encoded by $\mathbf{W}$, the county adjacency matrix, defined as
\begin{equation}
w_{kk}=0\, \hspace{3mm} \textrm{and}\, \hspace{3mm}
w_{kl}=\begin{cases}
1 & \textrm{if regions {\it k} and {\it l} are adjacent}\\
0 & \textrm{otherwise}
\end{cases}
\end{equation}
Here, $D = \mathrm{diag}\{g_1,g_2,\ldots,g_{R}\}$ where $g_i$ is the number of regions adjacent to region $i$.

The second term captures prediction-dependent, uncorrelated noise with additive and multiplicative components governed by $\sigma_a$ and $\sigma_b$, respectively. The relative contribution between the correlated GMRF noise and the uncorrelated noise is controlled by $\tau_{\Phi}$. Full details, including the development of $P$, can be found in the Part I paper\cite{24sr3a,safta2024detecting}. The predictive skill and flexibility of this epidemiological model, as applied to one areal unit, are described in detail in Safta~\etal\cite{Safta:2021} and Blonigan~\etal\cite{Blonigan:2021} and the extension to joint estimation, described in the Part I paper, achieves much the same predictive accuracy.

\subsection{Statistical Inference}
\label{sec:si}

The set of parameters $\param$ defining the likelihood in Eq.~\eqref{eq:likelihood} is given by 
\begin{equation}
    \param = \mathrm{vec}\left(\theta_i\right) = \mathrm{vec}\left(  
    \begin{bmatrix}
    \mathbf{m}_1 & \cdots & \mathbf{m}_{R} & \mathbf{\eta}
    \end{bmatrix}
    \right)
    \label{eq:params}
\end{equation}
where $\mathbf{\eta} =(\tau_{\Phi}, \lambda_{\Phi}, \sigma_a, \sigma_m)$ are the global noise parameters. Inference consists of forming the posterior distribution $p\left(\param \ \vert \  \Data  \right) = p\left(\Data  \ \vert \ \param \right) p\left( \param \right) / p \left(\Data  \right)$ over uncertain parameters $\param$. As the posterior is intractable, we instead look to approximate it using VI. Hence, the following sections describe how VI is  formulated carried out to approximate the posterior $p(\param \ \vert \ \Data )$ for the outbreak model as well as how the prior $p(\param)$ is defined to regularize the inverse problem.

\subsubsection{Variational Inference}
\label{sec:mfvi}

We will compare the Bayesian posterior sampled with AMCMC with posterior models obtained using MFVI which recasts approximate inference as an optimization problem. In particular, as the exact posterior is intractable, we consider a family of approximating densities $\mathcal{F}= \{q(\param;\boldsymbol{\phi}) \ \vert \  \boldsymbol{\phi} \in \boldsymbol{\Phi} \subseteq \mathbb{R}^d \}$ and seek to find a density $q(\param;\boldsymbol{\phi}^*)$ that minimizes the KL-divergence with respect to the posterior
\begin{equation}
    \boldsymbol{\phi}^* = \mathrm{argmin} \hspace{2mm} D_{\mathrm{KL}} \left( q(\param;\boldsymbol{\phi})\  \lVert \ p(\param \ \vert \  \Data ) \right)
\end{equation}
This can be re-expressed as minimizing the objective function  $\mathcal{L}(\boldsymbol{\phi})$ based on the evidence lower bound (ELBO) \cite{Kingma:2019}
\begin{equation}
    \mathcal{L}(\boldsymbol{\phi}) = -\mathbb{H} \left[q(\param;\boldsymbol{\phi}) \right] - \mathbb{E}_{q(\param;\boldsymbol{\phi}) } \left[\log p(\Data  \ \vert \  \param) + \log p(\param) \right]
    \label{eq:ELBO}
\end{equation}
where the first term in Eq.~\eqref{eq:ELBO} is the entropy of the surrogate posterior and the second, data-dependent term is an expectation with respect to the surrogate posterior that reflects both the expected data-fit and the prior. Here we take $\mathcal{F}$ to be the set of mean-field Gaussian distributions, i.e., 
\begin{equation}
    q\left(\param;\ \boldsymbol{\phi} \right) = \prod_{i=1}^d q_i\left(\theta_i; \ \mu_i, \sigma_i\right)
    \label{eq:mean_field_Gaussian}
\end{equation}
where $q_i(\theta_i ; \mu_i,\sigma_i) = \mathcal{N}(\theta_i ; \mu_i,\sigma_i)$, $\boldsymbol{\phi} = (\boldsymbol{\mu},\boldsymbol{\sigma})$ and we arrive at an optimization problem over $2d$ parameters where $d$ is the number of parameters defining the epidemiological model $\mathcal{M}$. To carry out the above minimization problem, we aim to use a gradient-based iterative scheme as the expectation in Eq.~\eqref{eq:ELBO} cannot be evaluated explicitly due to the nonlinearity of the forward model. Furthermore, $\mathcal{L}(\boldsymbol{\phi})$ is potentially a non-convex objective. Note that the gradient and expectation operators do not commute, i.e., 
\[
\nabla_{\boldsymbol{\phi}} \mathbb{E}_{q \left(\param;\boldsymbol{\phi} \right) } \left[ \log p \left(\Data  \ \vert \  \param \right) p(\param) \right] \neq \mathbb{E}_{q \left(\param;\boldsymbol{\phi} \right) } \left[\nabla_{\boldsymbol{\phi}} \log p \left(\Data  \ \vert \  \param \right) p \left(\param\right) \right]
\]
so some care has to be taken to arrive at a Monte Carlo estimator for the gradient $\nabla_{\boldsymbol{\phi}} \mathcal{L}(\boldsymbol{\phi})$. Two widely used approaches are: (a) the score function estimator, described in \S~\ref{sec:appendix-ELBO-score} (in the Appendix), which forms the basis of black-box VI and requires only evaluations of the log-likelihood, and (b) the reparametrization approach which requires gradients of the log-likelihood. The score function estimator typically displays much larger variance as seen in Kucukelbir \etal \cite{Kucukelbir:2017} where two orders of magnitude more samples were needed to arrive at the same variance as a reparametrization estimator. A similar trend was confirmed for the outbreak problem (Fig.~\ref{fig:ELBO_conv}) suggesting that the reparametrization approach would lead to superior scalability. Reparametrization proceeds by expressing $\param$ as a differentiable transformation $\param = t(\boldsymbol{\epsilon},\boldsymbol{\phi})$ of a $\boldsymbol{\phi}$-independent random variable $\boldsymbol{\epsilon} \sim q(\boldsymbol{\epsilon})$ such that $\param(\boldsymbol{\epsilon},\boldsymbol{\phi}) \sim q(\param, \boldsymbol{\phi})$. This allows the gradient to be expressed as
\begin{equation}
    \nabla_{\boldsymbol{\phi}} \mathcal{L} \left(\boldsymbol{\phi} \right) = - \nabla_{\boldsymbol{\phi}} \mathbb{H} \left[ q \left(\param;\boldsymbol{\phi} \right) \right] - \mathbb{E}_{q\left(\boldsymbol{\epsilon} \right) } \left[\nabla_{\boldsymbol{\phi}}  \log p \left(\Data  \ \vert \  \param(\boldsymbol{\epsilon},\boldsymbol{\phi}) \right) + \nabla_{\boldsymbol{\phi}} \log p \left(\param \left(\boldsymbol{\epsilon},\boldsymbol{\phi} \right) \right) \right]
    \label{eq:ELBO-grad}
\end{equation}
where gradients of the entropy term in Eq.~\eqref{eq:ELBO-grad} are available analytically for the Gaussian surrogate posterior and the second term can now be approximated with Monte Carlo given a method to compute the required gradients. For many machine learning models, automatic differentiation can be exploited to calculate the gradient of the log-likelihood with respect to parameters $\param$. Here, the objective function involves the log of the likelihood (Eq.~\eqref{eq:likelihood}) where derivatives of matrix inverses and determinants with respect to parameters are required to compute the gradient. Gradients such as these are not available using most automatic differentiation libraries. Instead, matrix calculus and quadrature were used to compute the derivatives of the log-likelihood with respect to model predictions $\mathbf{y}^{(p)}_i$ and to approximate the derivatives of the model predictions with respect to parameters, respectively. For details, see \S~\ref{sec:appendix-VI} (in the Appendix).

Note that some of the parameters comprising $\param$ are required to satisfy constraints for the noise and epidemiological models to be well-defined. For example, noise parameters $\tau_{\Phi}$, $\sigma_a$, and $\sigma_m$ as well as model parameters $N^r$, $k^r$ and $\theta^r$ for $r=1,\ldots,R$ should be positive while $\lambda_{\Phi}$ should satisfy $0 \leq \lambda_{\Phi} < 1$. Sampling from the mean-field Gaussian (Eq.~\eqref{eq:mean_field_Gaussian}) during the Monte Carlo estimation of the gradient (Eq.~\eqref{eq:ELBO-grad}) may result in violations of these constraints. To maintain the required properties without resorting to constrained optimization, we express a constrained parameter $\theta_i$ as an invertible, differentiable transformation $\theta_i = f_i(\hat{\theta}_i)$ of an unconstrained $\hat{\theta}_i$. Hence, the distribution governing $\theta_i$ is the push-forward density of $\mathcal{N}(\hat{\theta}_i ; \mu_i,\sigma_i)$ through $f_i$, i.e., the components of the mean-field surrogate posterior (Eq.~\eqref{eq:mean_field_Gaussian}) have modified probability densities
\begin{equation}
     q_i(\theta_i ; \mu_i,\sigma_i) = \mathcal{N}(\hat{\theta}_i ; \mu_i,\sigma_i) | f_i'(\hat{\theta}_i) |^ {-1}; \hspace{2mm} 1 \leq i \leq d
    \label{eq:mean_field_Gaussian_pf}
\end{equation}
where $\hat{\theta}_i=f_i^{-1}(\theta_i)$. This results in mean-field approximation where some of the factors are Gaussian and others non-Gaussian. Each factor is still defined by a $\mu_i$ and $\sigma_i$ parameter. The transformations are listed in Table~\ref{tab:variable-transformations}.

Note that MFVI has a tendency to underestimate the uncertainty / variance in the estimated parameters\cite{Han:2019} and in \S~\ref{sec:res}, we will check if this poses a limitation on the usefulness of the approximate solution, especially in the predictive skill of model $\mathcal{M}$ and therefore the detection of the start of outbreaks.

\subsubsection{Prior distribution}
\label{sec:prior}

The COVID-19 case count data exhibits significant noise due to inaccurate case counting reported by hospitals. Furthermore, counties with small populations exhibit sparse data in the sense that not many positive daily case counts were reported. Hence, we expect the inverse problem to be ill-posed and require  regularization in the form of a prior $p(\param)$ over the parameters. 

Because of push-forward formulation described by Eq.~\eqref{eq:mean_field_Gaussian_pf}, a number of the parameters are already constrained by transformations $\theta_i = f_i(\hat{\theta}_i)$. In particular, the parameters $N^r, k^r, \theta^r$ for $r=1,\ldots,R$ and each of the noise parameters comprising $\mathbf{\eta}$ are all constrained by transformations to take on values in some restricted interval. For example, $\lambda_{\Phi}$ is constrained to lie within $[0,1-\epsilon]$, for some $\epsilon >0$ so that Eq.~\eqref{eq:noise-model} defines a valid covariance matrix, i.e., it remains symmetric, positive definite. The parameters $t_0^r$ are the only unconstrained variables. Hence, we take Gaussian priors over $t_0^r$ that incorporate diffuse assumptions about when it is reasonable for a wave to occur.

\subsubsection{Posterior predictive tests}
\label{sec:ppt}

The mean-field variational inference (MFVI) described in \S~\ref{sec:mfvi} results in a multivariate Gaussian posterior distribution (Eq.~\eqref{eq:mean_field_Gaussian_pf}) that then needs to be verified against data $\Data$. To do so, we take samples $\param_j \sim q\left(\param; \boldsymbol{\phi} \right)$, and using Eq.~\eqref{eq:obs_eqn}, generate predictions $\Ypred_j = \{\mathbf{y}^{(p)}_i + \boldsymbol{\epsilon}_i\}_j, j = 1 \ldots J$. For this paper, $J = 100$. The time-series $\Ypred_j$ result in a ``fantail'' of predictions for each areal unit $r$ which should statistically reproduce $\Data$ e.g., the inter-quartile range of the samples $\Ypred_j$ should bound 50\% of the observations and the $5^{th}$ and $95^{th}$ percentiles should bound 90\% of the individual data points in $\Data$. This test is called a posterior predictive test (PPT) which we will use extensively in \S~\ref{sec:res}. In addition, we define a score to summarize the agreement of $\Ypred_j$ with $\Data$. Predictive distributions constituted out of $\{\mathbf{y}^{(p)}_i \}_j, j = 1 \ldots J$ samples are called ``push-forward`` (PF) predictions.

Define $y_{r, i}^{(pred)}(\param) = y_r(i; \ \paramDis_r) + \epsilon_r(\boldsymbol{\eta})$ to be the model prediction for region $r$ for day $i$ by combining Eq.~\eqref{eq:obs_eqn} and Eq.~\eqref{eq:params}. Let $y_{r,i,j} = y_{r, i}^{(pred)}(\param_j)$ be the prediction corresponding to a sample $\param_j$ drawn from the posterior. Let $y^{(obs)}_{r,i}$ be the corresponding observation (from $\Data$). Let $G_{r,i}(y^{(pred)}; \Theta)$ be the cumulative distribution function (CDF) for the model predictions arising from the posterior distribution $q(\Theta; \boldsymbol{\phi})$. The Continuous Ranked Predictive Score (CRPS~\cite{07gr2a}) is defined as
\begin{equation}
    c_{r,i} =   \int \left(G_{r,i}(\upsilon) - \mathbbm{1}_{\upsilon > y^{(obs)}_{r,i}} \right)^2 d \upsilon 
    \hspace{3mm} \mbox{and} \hspace{3mm}
    C_r  =  \frac{\sum^{N_d}_i c_{r,i}}{N_d}.
    \label{eq:crps}
\end{equation}
The CRPS has units of case-counts and will be larger for areal units with larger total case-counts $T_r = \sum_i y^{(obs)}_{r,i}$, and we will use the  ratio $\rho_r = C_r / T_r$ to compare across areal units in \S~\ref{sec:res}. In practice, the empirical CDF computed using $J$ samples of $\Ypred_j$ is used to approximate $G_{r,i}(y^{(pred)}; \Theta)$. This score function has been used in judge the quality of the model to capture the spread in the data used for calibration.~\cite{Safta:2015,Safta:2021,Hegde:2023} We use the implementation in the R Statistical Software\cite{Manual:R} (R version 4.3.2 (2023-10-31)) package \texttt{verification}\cite{16ss3a}, specifically the function \texttt{crpsDecomposition()}, to compute the CRPS.

\section{Results}
\label{sec:res}
\begin{figure}[h!]
    \centerline{
        \includegraphics[width = 0.5\textwidth]{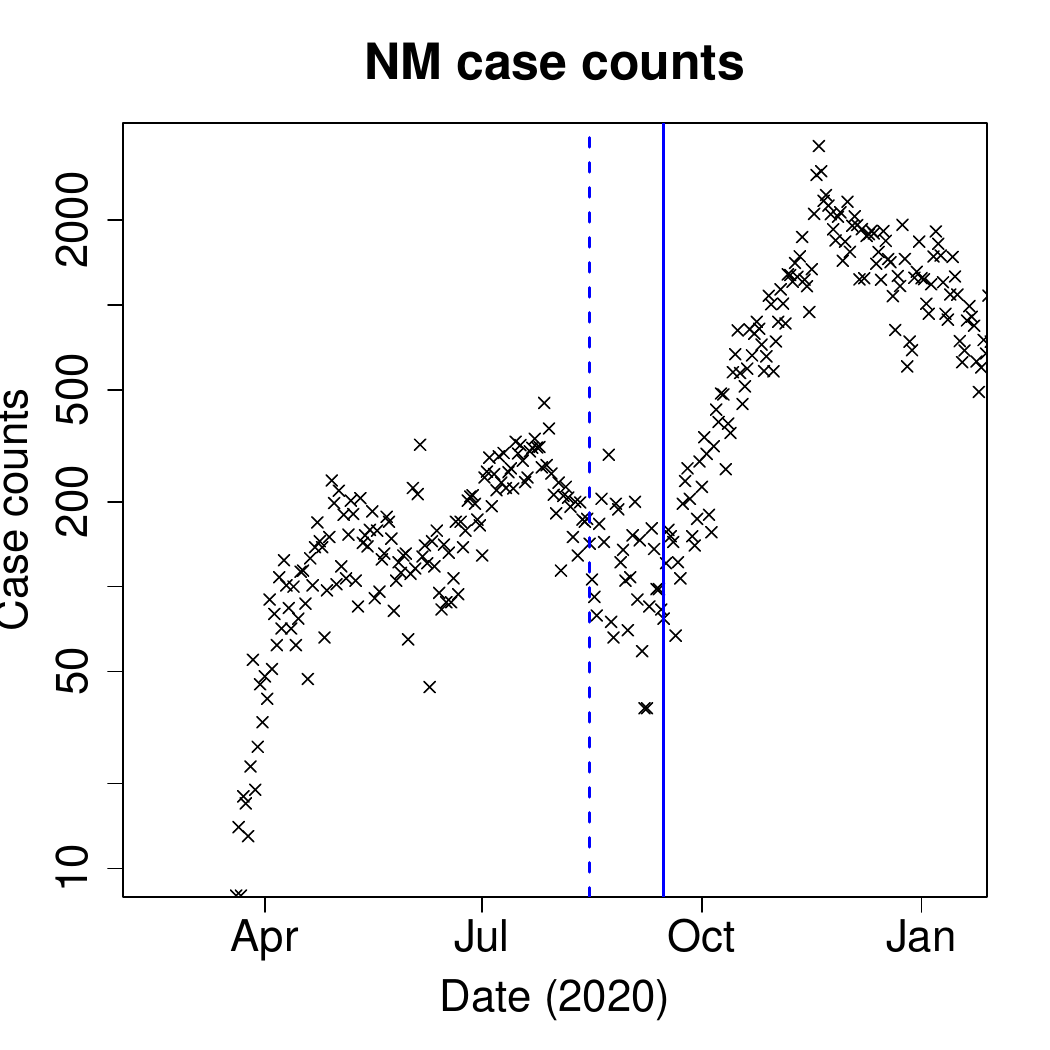}
        \includegraphics[width = 0.53\textwidth]{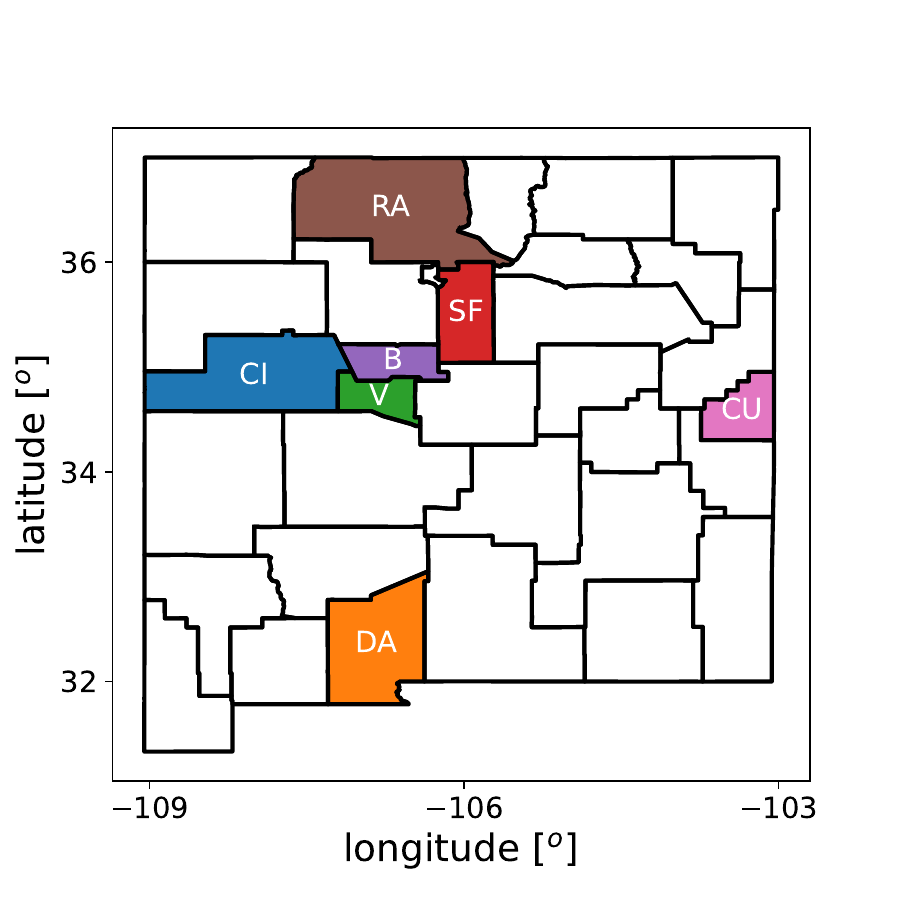}}
    \caption{Left: The three COVID-19 waves in NM in 2020. We see the Spring wave, ending around June $1^{\text{st}}$, followed by the Summer wave and the Fall wave that started around September $15^{\text{th}}$. Our aim is to estimate the infection-rate \emph{field} during the Summer wave and detect the Fall 2020 wave. The solid vertical line denotes September $15^{\text{th}}$ and the dashed one is placed at August $15^{\text{th}}$. Right: The couties of New Mexico. The shaded ones are where we will present results. Abbreviations: B = Bernalillo; CI = Cibola; CU = Curry; DA = Do\~{n}a Ana; RA = Rio Arriba SF = Santa Fe and V = Valencia.}
    \label{fig:FallWave}
\end{figure}

In this section, the calibration of the outbreak model, using the formulation presented in \S~\ref{sec:form}, is studied across several cases. The COVID-19 pandemic arrived in NM in March 2020; Fig.~\ref{fig:FallWave} (left) plots the detected cases in NM over 2020. We see three clear ``waves`` - the Spring wave, the Summer wave which spanned June $1^{\text{st}}$ to September $15^{\text{th}}$, and the Fall wave that arrived after that. Unless specified otherwise, data from the Summer wave (i.e., June 1, 2020 to September 15, 2020) is used to perform the estimation and the estimated infection-rate is used to forecast two weeks ahead, into the Fall wave. The COVID-19 dataset of case-counts that we use covers the duration from 2020-01-22 to 2022-05-13, and consists of daily (new) case-counts of COVID-19 from each of the 33 counties of NM; the data is available online.~\cite{nytcovid,jhucovid}. The sudden change in epidemiological dynamics around September $15^{\text{th}}$ implies that that the infection-rate profile from the Summer wave will not be able to forecast the Fall wave; therefore any disagreement between the observed data and forecasts is a sign of the arrival of the Fall wave.

First, we consider three populated counties that are adjacent to each other and display a large number of cases. These counties present less noise and clearer trends resulting in a more well-posed inversion task. The aim of the study is to compare the effect of calibrating multiple areal units jointly using MFVI, as well as to compare against the calibration performed using AMCMC in our Part I paper\cite{24sr3a,safta2024detecting}. Next, the coupled outbreak model is calibrated across all 33 New Mexico counties. This represents a more challenging task given the dimensionality of the problem as well as a multitude of counties displaying sparse and noisy case-counts.  We then evaluate the final posterior approximation using PPT runs.  The convergence of the MVFI procedure for both the outbreak model and noise parameters is investigated and discussed last. We next discuss anomaly detection using the MFVI-calibrated outbreak model is carried out using COVID-19 spread-rates across all 33 counties in New Mexico using data from the summer of 2020 to detect the arrival of the Fall 2020 COVID-19 wave.

When estimating the infection-rate field in the 3-counties, we infer $3 \times 4 + 4 = 16$ parameters -- 4 parameters $(t_0^r, N_r, k^r, \theta^r)$ for each county and 4 noise parameters $(\tau_\phi, \lambda, \sigma_a, \sigma_m)$; this was performed with AMCMC in our Part I paper. For all of NM, with its 33 counties, the dimensionality of the inverse problem is $33 \times 4 +4 = 136$ independent parameters. This is too large for AMCMC and was the motivation for developing MFVI.

The MFVI procedure follows \S~\ref{sec:form} where the stochastic gradient descent iteration is carried out using the Adaptive Moment Estimation (ADAM) algorithm \cite{Kingma:2014}. MFVI is well-known to display mode-seeking behavior due properties of the KL-divergence. Hence, to facilitate the convergence of MFVI, we initialize the mean parameters for MFVI from a Maximum Likelihood Estimate (MLE) which is readily available using gradients of the log-likelihood. During calibration via stochastic gradient descent, $N_s = 200$ samples were used in the Monte Carlo estimates of the ELBO gradient Eq.~\eqref{eq:ELBO-grad}.

\subsection{Three-county inversion}
\label{sec:3_county}

In this section we investigate the effect of using an approximate MFVI inversion by comparing against AMCMC solution computed in our Part I paper. We also check the effect of estimating the infection-rate across multiple areal units vis-\`{a}-vis independently.

\begin{figure}
    \centering
    \includegraphics[width=\textwidth]{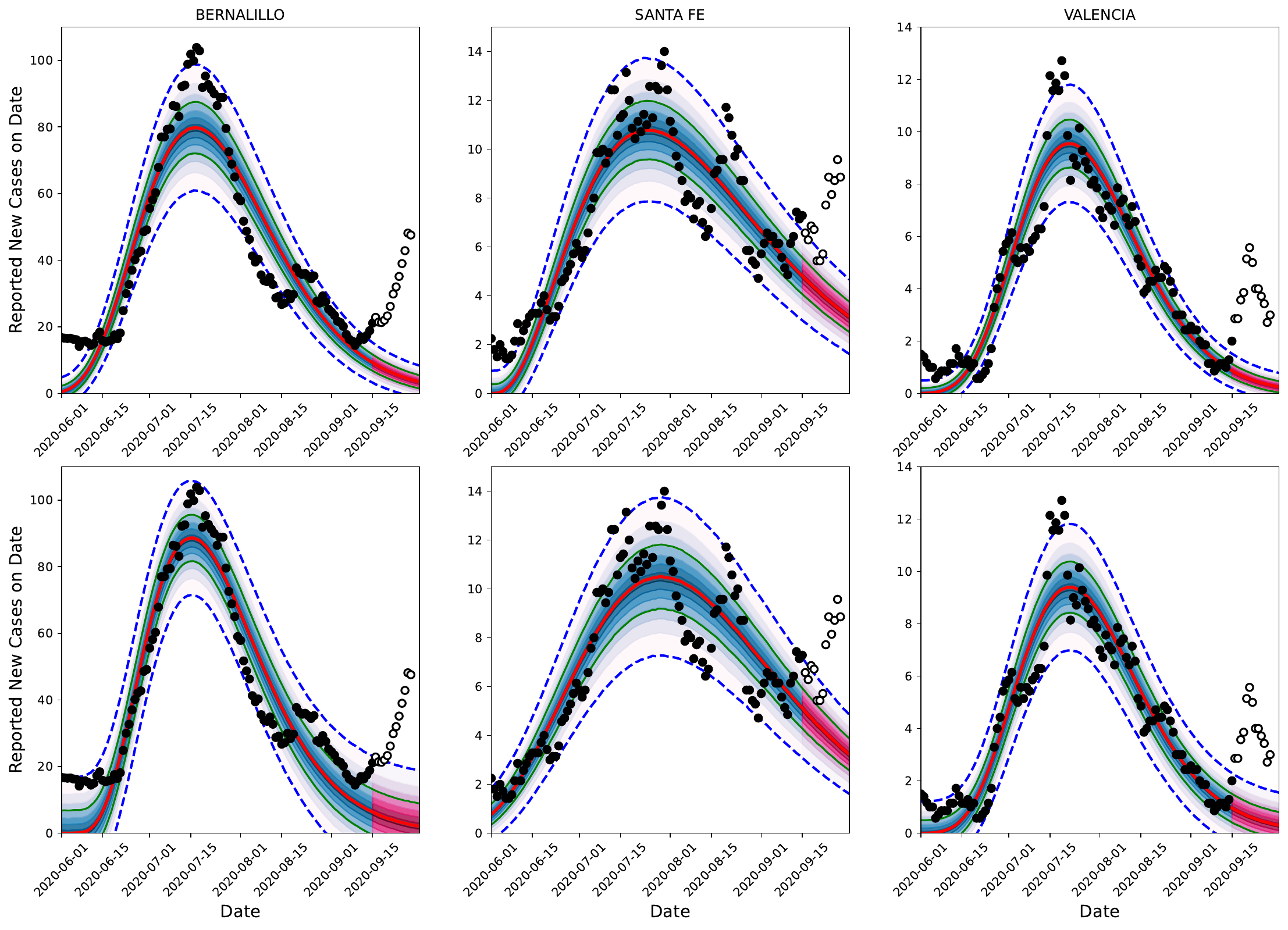}
    \caption{Comparison of the predictive distribution for the MFVI inversion of Bernalillo (left), Santa Fe (middle), and Valencia (right) done jointly using the \emph{GMRF} model (top) and \emph{independently} for each county (bottom). Case-count data was smoothed with a 7-day running average. The red line shows the median predictions, the shaded green line shows the inter-quartile range, and the dashed lines are the $5^{\text{th}}$ and $95^{\text{th}}$ percentiles. The filled symbols are the data $\Data$ used in the inversion. The unfilled symbols are the observations beyond September $15^{\text{th}}$ and are used to compare with the two-week forecasts.}
    \label{fig:3_county_pp}
\end{figure}
{\bf Joint versus independent calibrations:} First, we perform an infection-rate estimation for three adjoining counties - Bernalillo, Santa Fe and Valencia  - using MFVI  independently and jointly using the spatial (GMRF) model (Eq.~\eqref{eq:noise-model}); see Fig.~\ref{fig:FallWave} (right) for their positions.  In Fig.~\ref{fig:3_county_pp}, the results of a PPT for all three counties are displayed in both the joint and independently calibrated cases. The case-count data were smoothed with a 7-day running average. The median prediction is plotted with the red line and the dashed lines denote the $5^{\text{th}}$ and $95^{\text{th}}$ percentiles. We see that most of the observations (filled symbols) are within these bounds. We also see the 2-week-ahead forecast beyond September $15{\text{th}}$ and the unfilled symbols showing the observed data from that period. The forecasts and the data do not match for any of the three counties, indicating a change in the epidemiological dynamics. This change is due to the arrival of the Fall 2020 wave of COVID-19 in NM, and the figure shows that the wave arrived approximately simultaneously in all three counties (see Fig.~\ref{fig:FallWave} (left) for a clearer picture of the three COVID-19 waves encountered in 2020 in NM). Observe that the jointly calibrated predictive distribution displays less uncertainty in the predictions (i.e., small $\sigma_m$)  than the independently calibrated version. This is likely due to a combination of two factors. The first is that incorporated spatial correlations between counties regularizes the calibration and results in more certainty about the true underlying model parameters. The second is that the uncorrelated MFVI approximation is known to underestimate uncertainty for highly correlated distributions. Note also that uncertainty is largest in the predictive distribution for Santa Fe consistent with the markedly noisier behavior of the case-counts for this county. The infection-rate profiles for these counties are in Fig.~\ref{fig:3_county_fGamma_jo_in} in the Appendix. There is not much difference between them indicating that the infection-rate parameter estimates in the two cases might be similar, whereas the estimate of the noise, which affects PPT results, vary between the two formulations due to the fact that the spatial noise model has to accommodate all the noise, across all the areal units, together. These findings echo what we observed in our Part I paper, where the same study was performed using AMCMC as the estimation procedure.

\begin{figure}
    \centering
    \includegraphics[width = \textwidth]{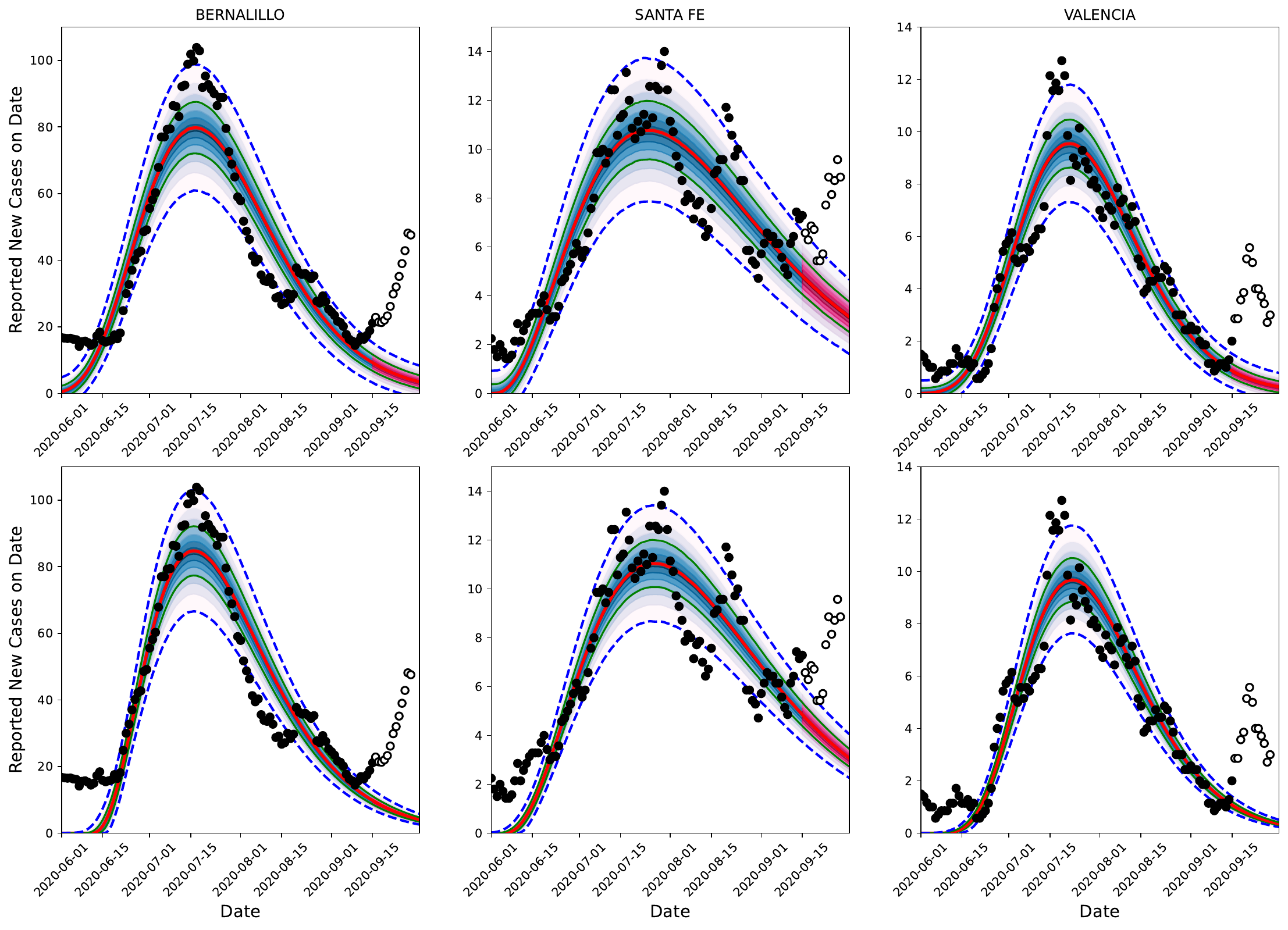}
    \caption{Comparison of the predictive distribution for the joint inversion of Bernalillo (left), Santa Fe (middle), and Valencia (right) done jointly using \emph{MFVI} (top) and \emph{AMCMC} (bottom). Case-count data was smoothed with a 7-day running average. The red line shows the median predictions, the shaded green line shows the inter-quartile range, and the dashed lines are the $5^{\text{th}}$ and $95^{\text{th}}$ percentiles. The filled symbols are the data $\Data$ used in the inversion. The unfilled symbols are the observations beyond September $15^{\text{th}}$ and are used to compare with the two-week forecasts. The AMCMC results are taken from our Part I paper\cite{24sr3a,safta2024detecting}.}
    \label{fig:3_county_pp_vimcmc}    
\end{figure}
{\bf AMCMC versus MFVI estimation:} Next, we study the effect of using our approximate MFVI method versus the estimates computed using AMCMC in the Part I paper. In Fig.~\ref{fig:3_county_pp_vimcmc} we plot the PPT results from the joint MFVI (top) and AMCMC (bottom) for the same three counties.  Here we see that the MFVI estimates a larger $\sigma_m$ - the $5^{\text{th}}$ and $95^{\text{th}}$ percentile bounds (dashed lines) are far wider for MFVI results vis-\`{a}-vis AMCMC results below. By dint of having wider bounds, the MFVI estimate is also better at bounding the data used to compute the infection-rate \emph{field} for the three counties. Again the arrival of the Fall 2020 wave is clearly discerned in the figure. The infection-rate profiles for these counties are in Fig.~\ref{fig:3_county_fGamma_vi_mcmc} in the Appendix and there is not much difference between them.

\begin{figure}
    \centering
    \includegraphics[width=1.0\textwidth]{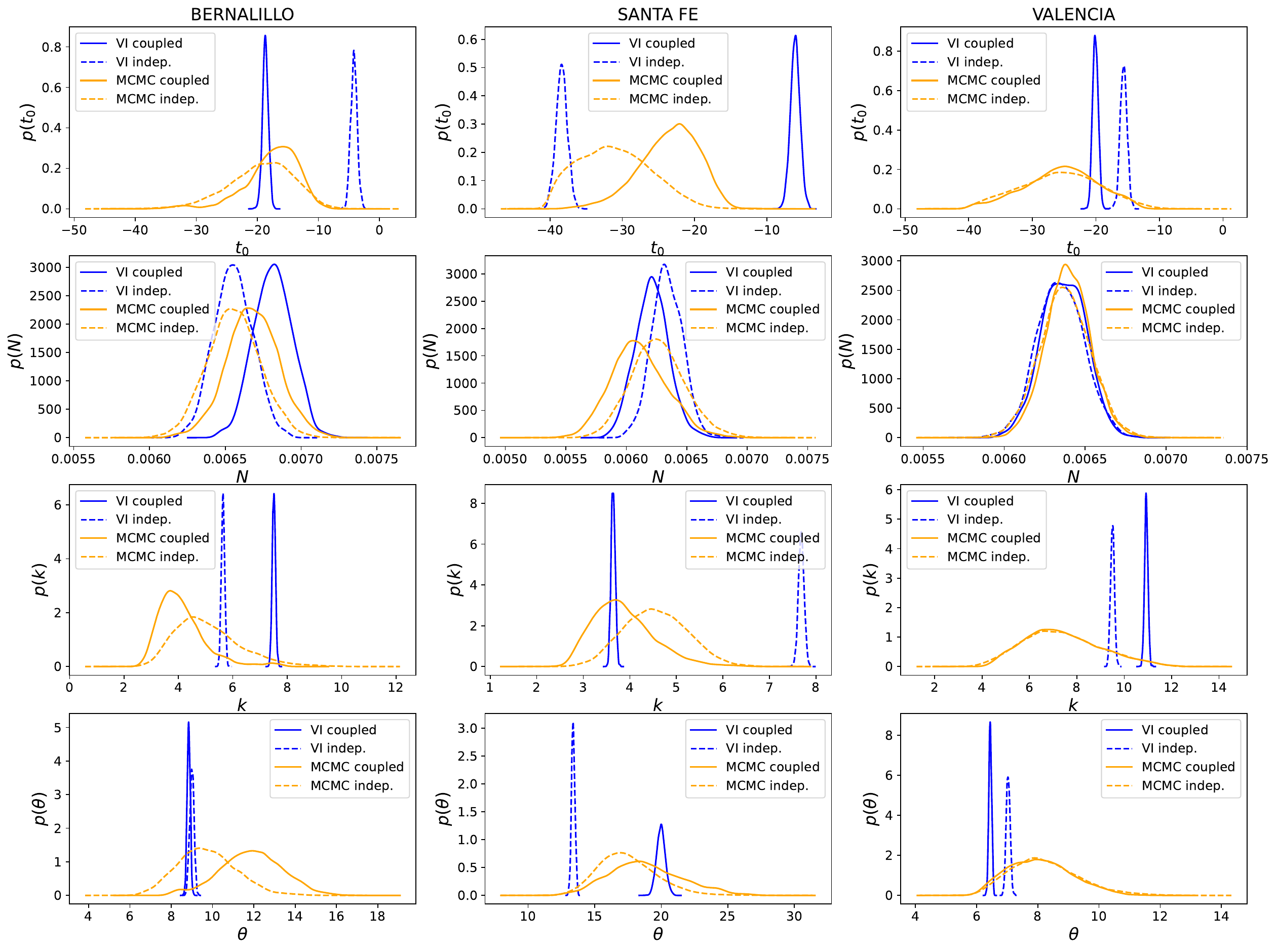}
    \caption{Comparison of the posterior over model parameters $t_0$, $N$, $k$, and $\theta$ for 3-county inference of Bernalillo (left), Santa Fe (middle), and Valencia (right) using both MCMC (orange) and MFVI (blue). Both joint (solid lines) and independent calibration (dashed lines) are results are displayed. $t_0$ values are negative as it is measured from June $10^{\text{th}}$, 2020, and the PDFs imply that infections for the Summer wave started in late May. 'VI indep.' in the legend implies an estimate obtained for each county independently using MFVI.}
    \label{fig:3_county_param_posterior}
\end{figure}
In Fig.~\ref{fig:3_county_param_posterior}, we summarize the marginalized posteriors for the infection-rate parameters for Bernalillo, Santa Fe, and Valencia computed using MFVI and AMCMC, jointly and independently. $t_0^r$ assumes negative values as it is measured from June $10^{\text{th}}$, 2020; the PDFs peak around -20 (for AMCMC results) and imply that the infections for the Summer wave started in late May, about 20 days before June $10^{\text{th}}$. The MFVI results also agree approximately with this estimation. We see that apart from $N$, the MFVI and AMCMC posteriors do not match. The MFVI posteriors are extremely narrow providing a spurious degree of certainty in the estimates; this arises from the form of the posterior distribution - independent Gaussians - that we postulate in Eq.~\ref{eq:mean_field_Gaussian}. In addition, as is clear from the AMCMC results, the ``true'' posteriors are not Gaussian. The marginalized posteriors computed via AMCMC, for joint and independent estimation, do match (sometimes very well, as in the case of Valencia), but they are wide apart for the MFVI, showing the effect of the Gaussian approximation. Yet the infection-rate profiles in Fig.~\ref{fig:3_county_fGamma_jo_in} (in the Appendix) do not show much of a difference, nor do the PPT results in Fig.~\ref{fig:3_county_pp}, leading us to conjecture that the influence of some of these parameter on the infection-rate may be muted. This can also be surmised from the marginal distributions computed from AMCMC - they are quite wide.

Finally, we summarize the predictive skill of the joint estimates computed using MFVI and AMCMC in Table~\ref{tab:crps} using CRPS i.e. $C_r$. The CRPS is computed over data between June $1^{\text{st}}$ and September $15^{\text{th}}$, 2020 i.e., the CRPS $C_r$ summarizes the agreement of the PPTs with observations for each of the counties. We see that the CRPS (a measure of the error between predictions and data) is about 10 cases per day, for Bernalillo (where case-counts peaked at about 100 cases/day; see Fig.~\ref{fig:3_county_pp}) and 2.5 cases a day for Valencia and Santa Fe (which peaked at about 12 cases a day). What is remarkable is the difference in the CRPS as computed using MFVI and AMCMC - it is less than a case-count per day. This small change in the predictive skill leads us to believe that the scalable, but approximate, MFVI approach might be sufficiently accurate to allow us to detect the arrival of the Fall 2020 wave in an automated manner.

\emph{To summarize}, joint estimation may lead to more certain forecasts (narrower PPTs) vis-\`{a}-vis independent estimation of infection-rates in  areal units (counties). MFVI provides spuriously low levels of uncertainty in estimated parameters. The marginalized posteriors of infection-rate parameters obtained from MFVI and AMCMC do not agree, but the infection-rate does not seem to be very sensitive to the disagreement. Consequently, the PPTs from the AMCMC and MFVI inversion are very similar in terms of their predictive skill.
\begin{center}
    \begin{table*}[!h]%
    \caption{Predictive skill of PPTs generated using infection-rate estimates computed using different procedures. All estimations are performed using the joint formulation, using the spatial model. The PPTs are scored using CRPS. They have units of ``case-counts''.}
    \label{tab:crps}
        \begin{tabular*}{\textwidth}{@{\extracolsep\fill}llll@{}} \toprule
        Procedure                         & Bernalillo & Santa Fe   & Valencia    \\ \midrule
        AMCMC, 3-county$^{\tnote{\bf a}}$  & 11.3       & 2.65       &  1.76  \\
        MFVI, 3-county$^{\tnote{\bf b}}$  & 10.75      & 2.54       &  1.70  \\
        MFVI, 33-county                   & 10.7       & 2.35       &  1.47  \\ \bottomrule
    \end{tabular*}
    \begin{tablenotes}
    \item[$^{\rm a}$] Taken from Table 2 of our Part I paper\cite{24sr3a,safta2024detecting}.
    \item[$^{\rm b}$] Computed from Fig.~\ref{fig:3_county_pp} (top).
    \end{tablenotes}
    \end{table*}
\end{center}

\subsection{Joint inversion of all NM counties}
\begin{figure}
    \centering
    \includegraphics[width = 0.5\textwidth]{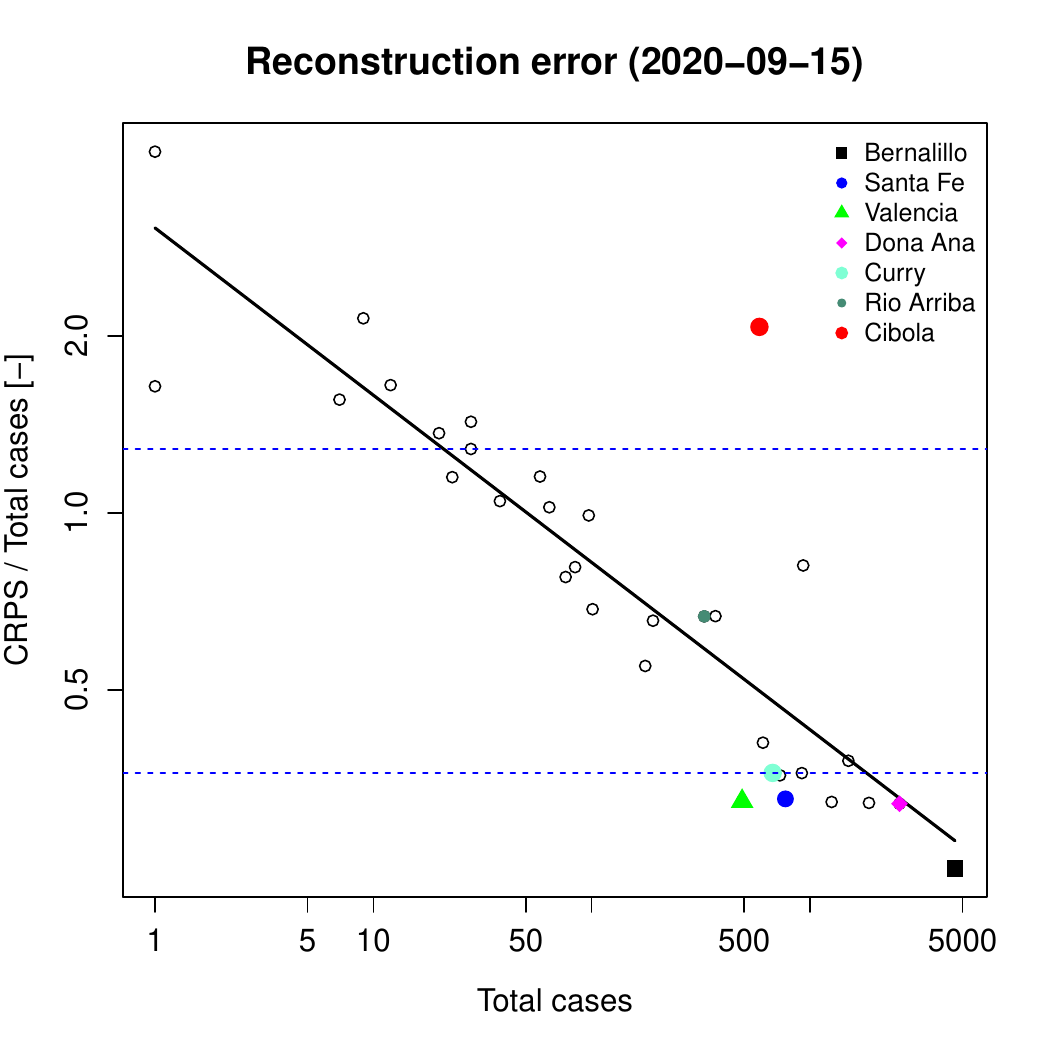}
    \caption{$C_r/T_r$ plotted as a function of $T_r$. Both the axes are log-axes. We see, as expected, that the predictive skill of the disease model, post calibration, is better for counties with larger case-counts $T_r$ where noise variance is low. The horizontal dashed lines are the first and third quartiles of $\rho = C/T$.}
    \label{fig:acc}
\end{figure}
Next, calibration of the full 33-county model using MFVI was carried out, followed by PPT runs which were then summarized using CRPS i.e., $C_r$ was computed for each NM county. In Fig.~\ref{fig:acc}, we plot the CRPS normalized by the total number of cases i.e., $C_r/T_r$ as a function of $T_r$, for $r = 1,\ldots,R$. Data between June $1^{\text{st}}$ and September $15^{\text{th}}$ was used in the infection-rate estimation as well as the computation of CRPS i.e., $C_r$ is a measure of the "goodness of model fit" to data. We see that the CRPS, as a ratio of the total cases, decreases with increasing number of total cases, as the disease model fits larger outbreaks in counties like Bernalillo. Others, like Cibola, do not agree with model predictions, due to flaws in the data, as we will see later in Fig.~\ref{fig:tempDetect33}. The straight line fit to data has the form
\[
    \log(\rho) = \log \left( \frac{C}{T} \right) \approx 1.1 - 0.28 \log (T),
\]
where $C$ and $T$ are the CRPS and total number of cases. This scaling shows that $\rho$ scales (approximately) as the fourth root of the total number of cases, a slow reduction indeed.

The horizontal lines in Fig.~\ref{fig:acc} show the first and third quartiles of $\rho$. The three counties studied in our Part I paper, Bernalillo, Santa Fe and Valencia, are marked and fall in the lowest quartile i.e., their data is good and the calibrated disease model is predictive. We will use the county of Do\~{n}a Ana as another member of the ``good'' class, while Curry and Rio Arriba, which fall in the inter-quartile range of $\rho$, will serve as exemplars of the ``middling'' class of calibration. Cibola, which falls in the last quartile, will be an exemplar of the ``bad'' class of calibration. We now examine the quality of the inversion in each of these counties.

\begin{figure}
    \centerline{
        \includegraphics[width = 0.4\textwidth]{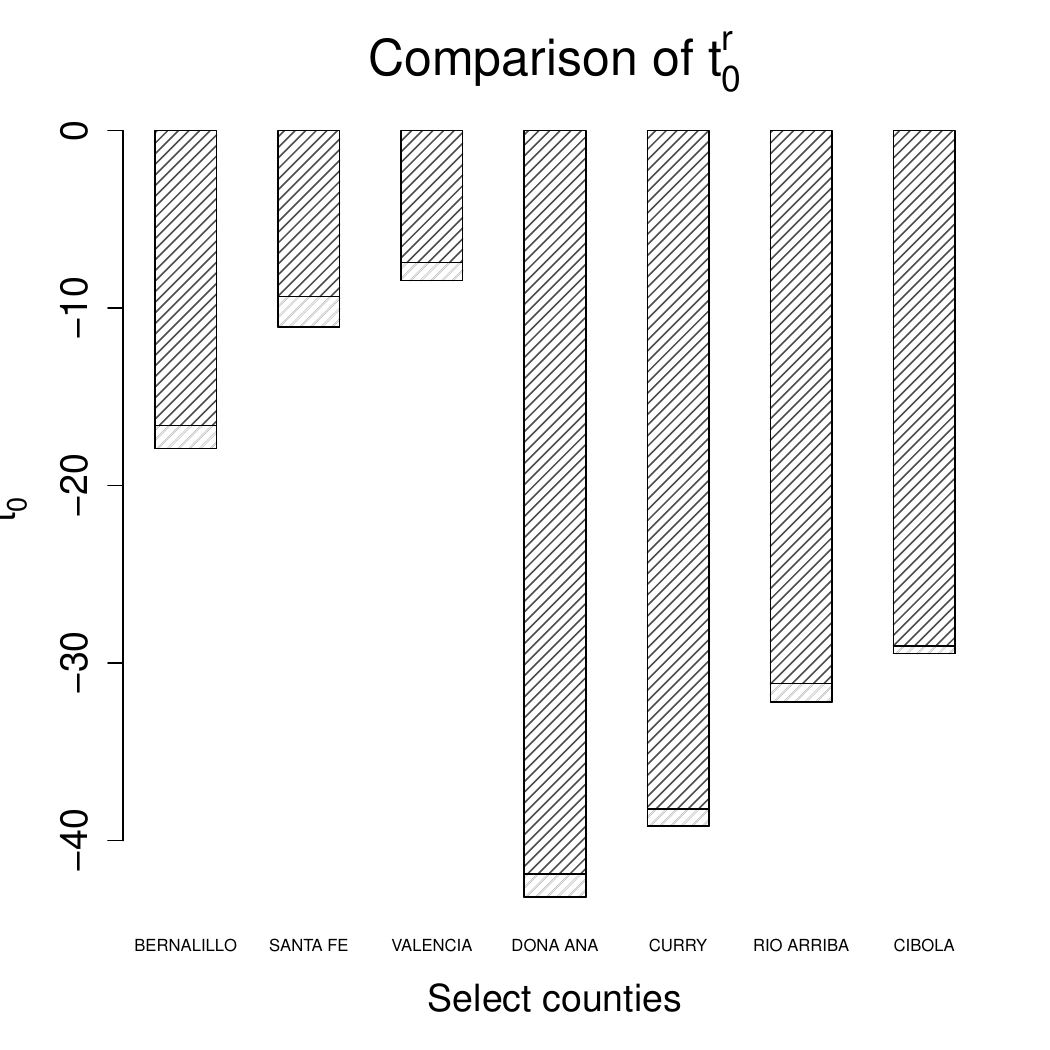}
        \includegraphics[width = 0.4\textwidth]{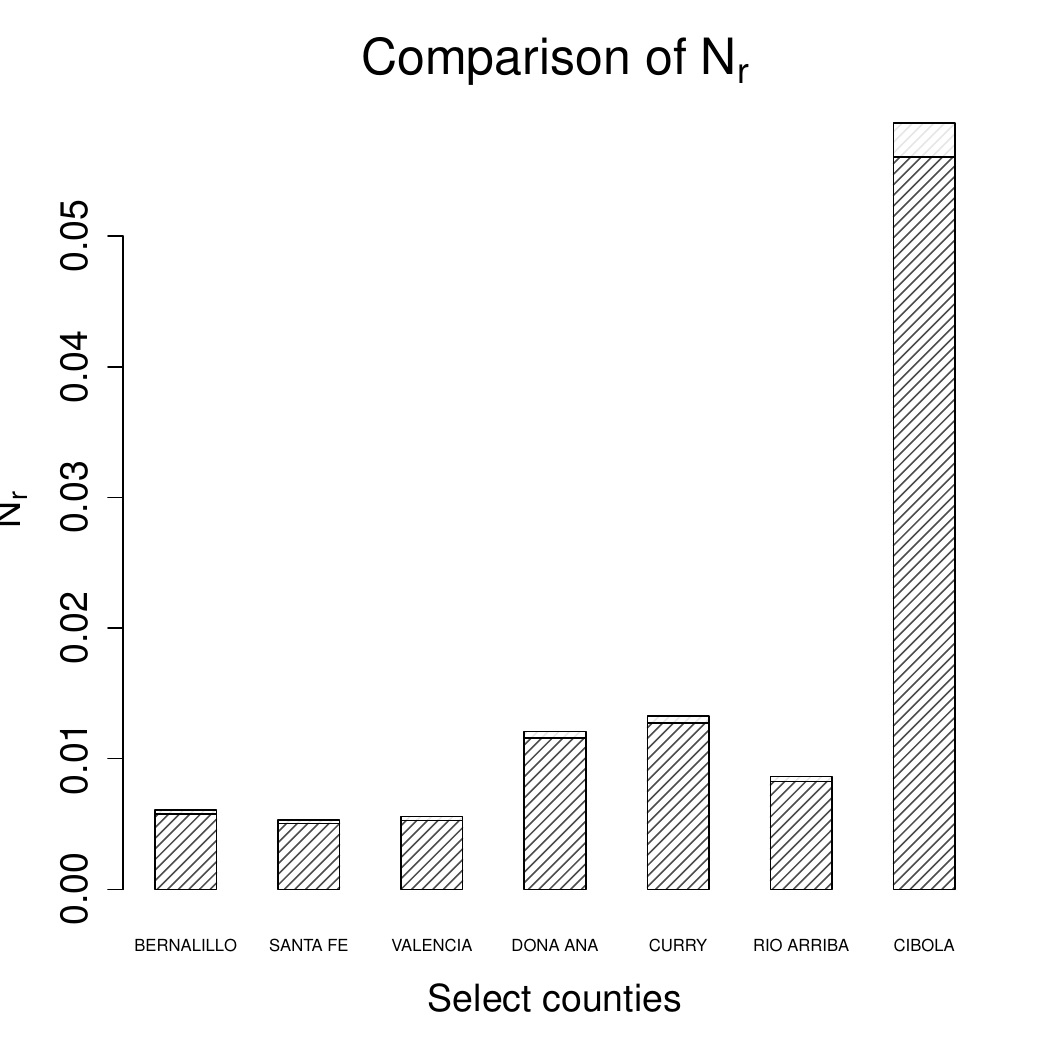}}
    \centerline{
        \includegraphics[width = 0.4\textwidth]{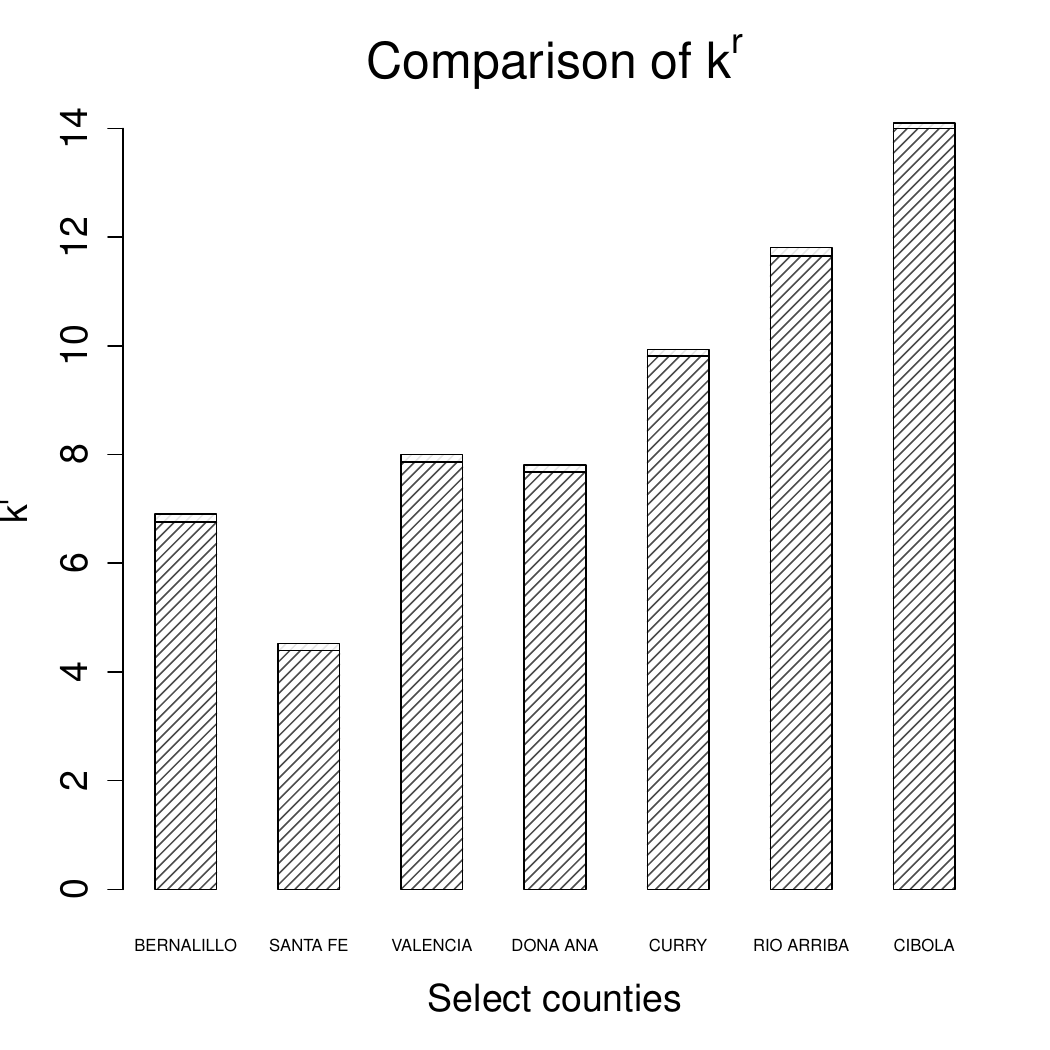}
        \includegraphics[width = 0.4\textwidth]{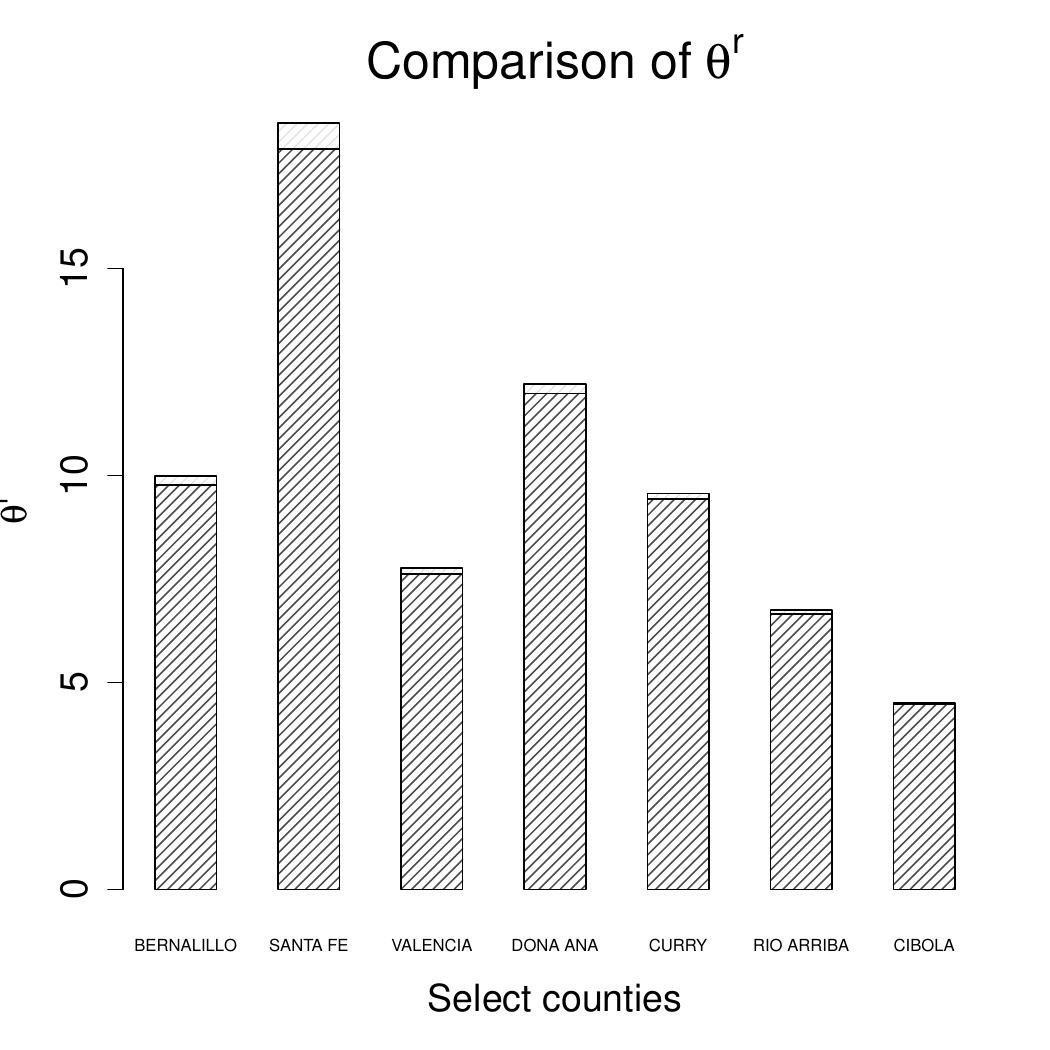}}
    \caption{Means (dark) and standard deviation (gray) of various infection-rate parameters, for select counties marked in Fig.~\ref{fig:acc}. Top left: The start time of the Summer wave $t^r_0$. Top right: The total size of the Summer outbreak $N_r$. Bottom left: $k^r$, the shape parameter of the Gamma profile of the infection-rate. Bottom right: $\theta^r$, the scale parameter of the Gamma profile.}
    \label{fig:inv33}
\end{figure}
In Fig.~\ref{fig:inv33} we plot the infection-rate parameters, as estimated from the Summer wave case-count data, for 7 counties marked in Fig.~\ref{fig:acc}. We see that the standard deviations are spuriously tiny, in line with what was observed in Fig.~\ref{fig:3_county_param_posterior}; thus MFVI consistently underestimates the uncertainty in the parameter estimates. Further, this spuriously low uncertainty is pervasive - counties with high CRPS such as Cibola and Rio Arriba show much the same estimation uncertainties as counties such as Bernalillo and Santa Fe with CRPSs a factor of three smaller. Thus the uncertainty in the parameters' estimates do not seem credible and we will omit them from further discussion. However, despite the quality of the case-count data, the inversion completes \emph{stably} and provides plausible results. However, as Fig.~\ref{fig:acc} shows, the infection-rate may not be estimated very accurately in some counties and this might hamper the task of detecting the Fall 2020 wave reliably.

\begin{figure}
    \centerline{\includegraphics[width = 0.8\textwidth]{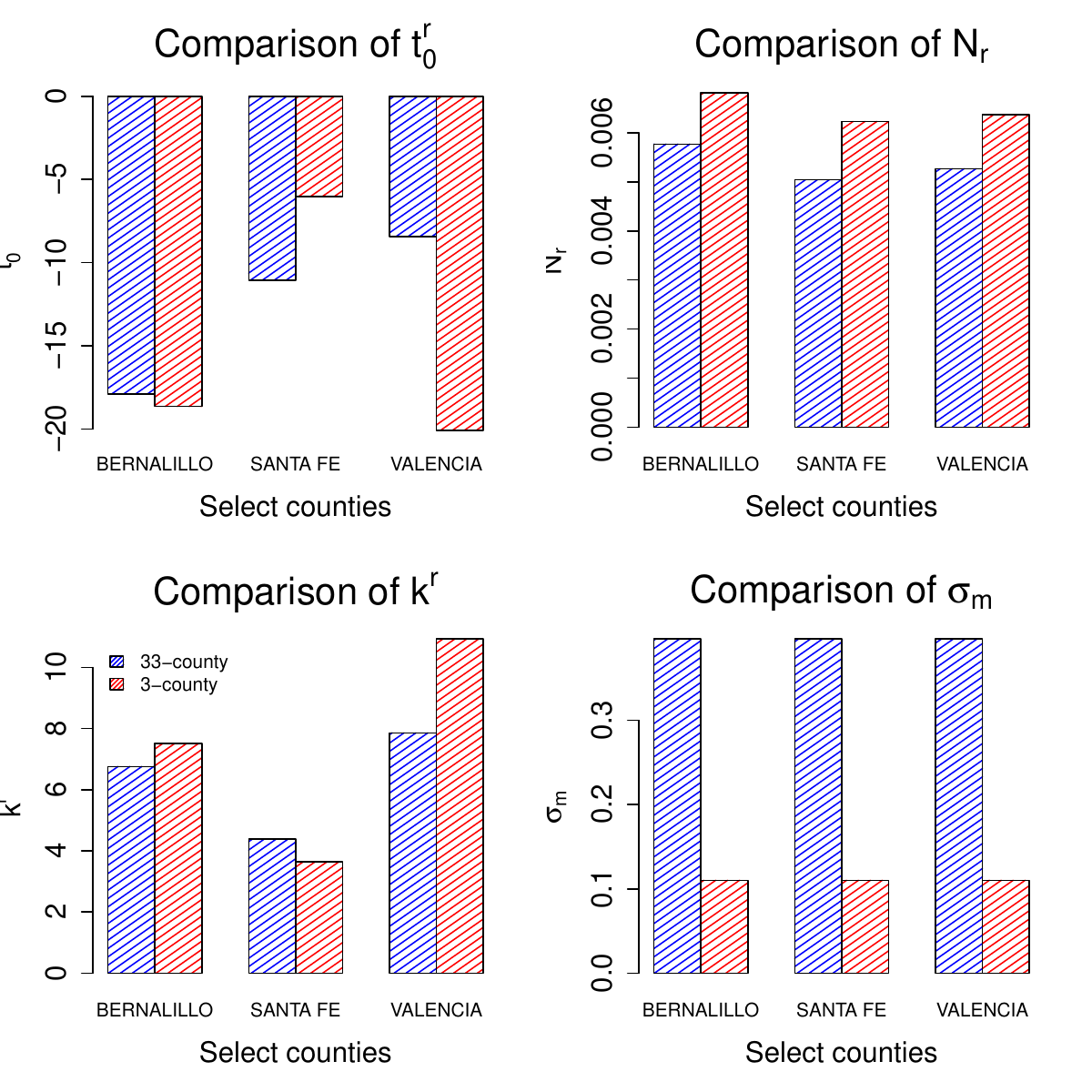}}
    \caption{Some infection-rate and noise parameters, estimated jointly among three counties, compared with their counterparts from a 33-county inversion. The noise parameter $\sigma_m$ is markedly larger in the 33-county inversion.}
    \label{fig:33comp03}
\end{figure}
In Fig.~\ref{fig:33comp03} we compare some infection-rate parameters $(t^r_0, N_r, k^r, \sigma_m)$ for Bernalillo, Santa Fe and Valencia, in the 3-county (as plotted in Fig.~\ref{fig:3_county_param_posterior}) and 33-county inversions. MFVI was used for both the computations.  We see that the parameters are not the same and there does not seem to be a clear trend, except that $\sigma_m$ increases when we include counties (some with poor quality data) in the estimation. This implies that the PPTs for the ``good'' counties will likely be wider than what could be achieved with AMCMC and this might have repercussions regarding detecting the Fall 2020 wave. The use of the AMCMC results to detect the Fall 2020 wave is described in our Part I paper - the wave was detected within a week of its arrival. 

\emph{To summarize}, the uncertainty in the estimated parameters, computed using MFVI, are not very credible. However, the MFVI inversion is stable, though the uncertainty in the PPTs will be larger because of an inflated $\sigma_m$, required to accommodate counties with very noisy data. 

\subsection{Algorithmic results}
\begin{figure}
	\centering
	\includegraphics[width=0.75\textwidth]{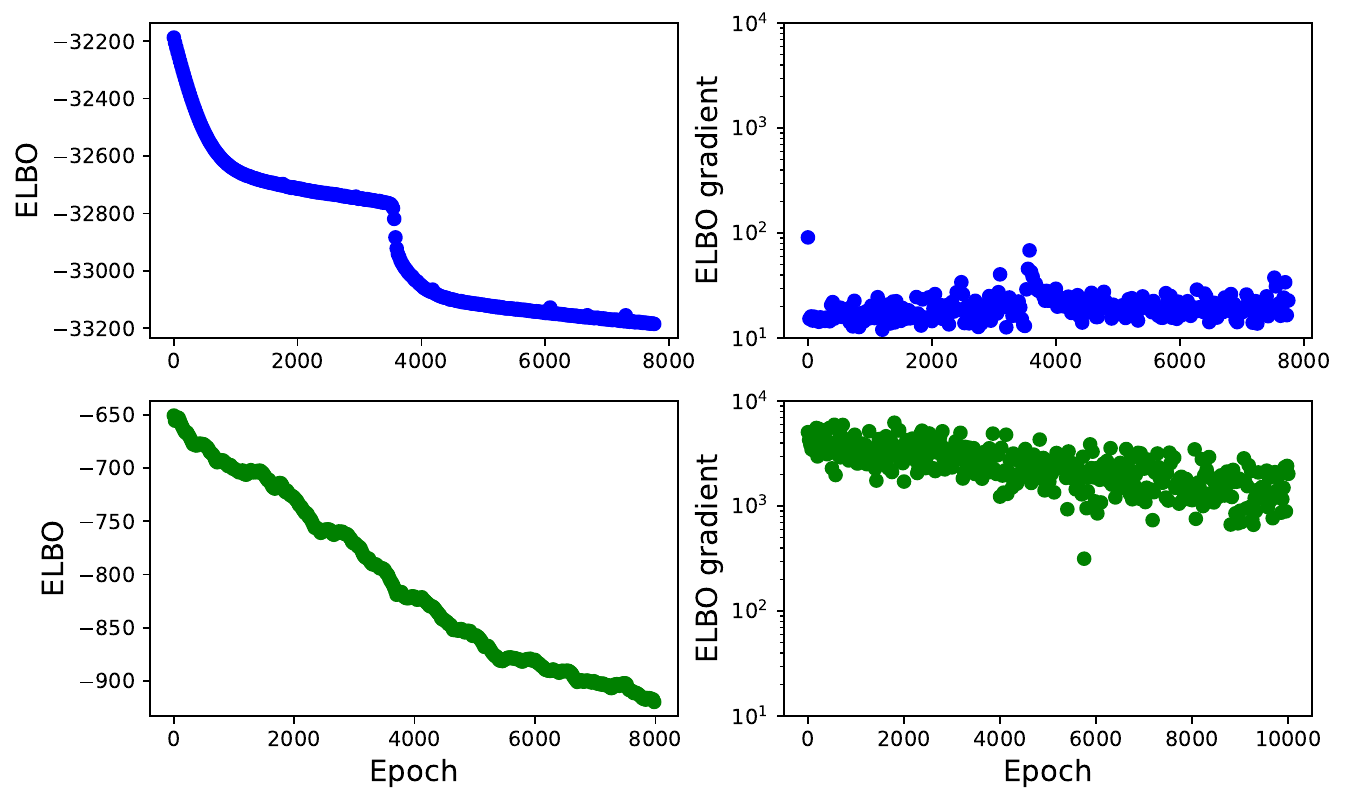}
	\caption{(Top) Convergence of the ELBO for the 33-county inversion along with the norm of the ELBO gradient as a function of gradient descent iterations for the reparametrized gradient formulation of MFVI. (Bottom) Convergence of the ELBO for a 1-county inverse problem along with the norm of the ELBO gradient for the black box formulation of MFVI. In both cases, $N_s=300$ samples were used for the MC estimators of the gradient.}
	\label{fig:ELBO_conv}
\end{figure}

Finally, we investigate the numerical aspects of the reparametrized algorithm described in \S~\ref{sec:mfvi}. The convergence of MFVI is depicted in Fig.~\ref{fig:ELBO_conv} where the ELBO and the norm of its gradient are shown as a function of gradient descent iterations.

The top of Fig.~\ref{fig:ELBO_conv} shows the ELBO and gradient for the 33-county calibration using the reparametrization formulation while the bottom provides a comparison to using black box VI for 1-county calibration of Bernalillo. In both cases, $N_s = 300$ samples were used for the reparametrization and score function MC estimators of the gradient. Note that even for a single county, black box VI shows significantly higher variance in the gradient leading to poor convergence in comparison to the much larger 33-county problem calibrated with reparametrization. The convergence of the ELBO in the top row is also quite smooth suggesting that significantly less samples could be used to obtain good estimates of the gradient with the reparametrization approach. Hence, it is clear that reparametrization is necessary to scale the calibration to the 33-county inversion despite the added complexity of obtaining gradients of the log-likelihood.

\begin{figure}
	\centering
	\includegraphics[width=0.8\textwidth]{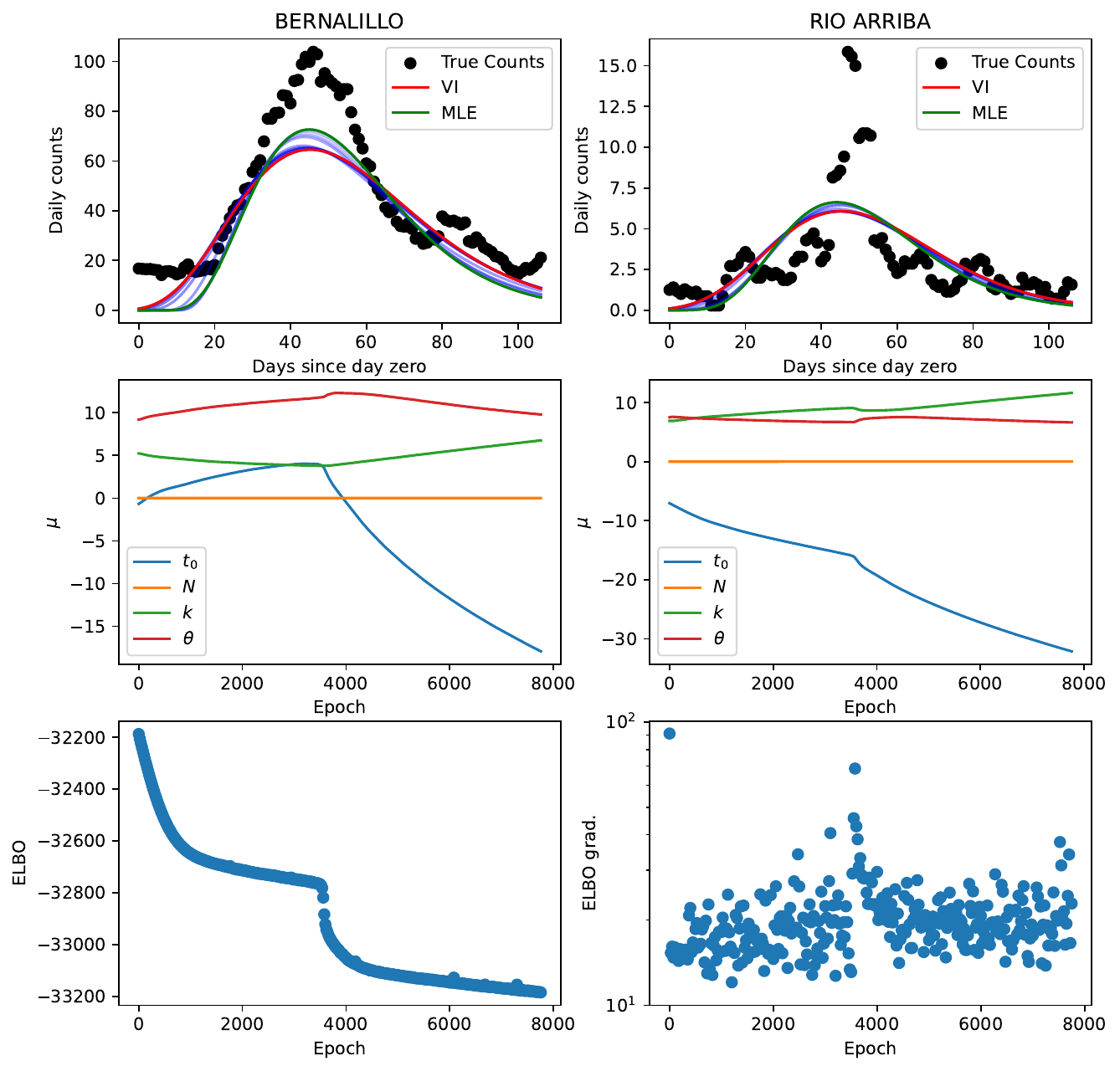}
	\caption{(Top) Visualization of the convergence of 33-county inversion using MFVI in terms of model predictions for two counties: Bernalillo (left) and Rio Arriba (right). The mean of the initial condition for MFVI is given by a MLE solution shown in red. Intermediate solutions are shown in blue along with the final mean solution in green. (Middle) Corresponding convergence of the model parameters for Bernalillo and Rio Arriba. (Bottom) The ELBO objective (left) and its gradient (right) as a function of iteration for the full 33 county inversion. Here, day zero of the inversion is defined as June $1^{\text{st}}$, 2020.}
	\label{fig:VI_conv}
\end{figure}

The top two rows of Fig.~\ref{fig:VI_conv} display convergence information from two counties, Bernalillo and Rio Arriba, taken from the 33-county inversion. Bernallilo and Rio Arriba were chosen as they have larger and smaller populations, respectively. The mean of the initial condition for MFVI is given by a MLE solution shown in red. Intermediate solutions are shown in blue along with the final solution in green. Observe that while similar to the MLE, MFVI subtly expands the shape of the wave to better cover the tail of the outbreak. This is potentially an effect of the tendency of the KL-divergence to increase the overlap of the surrogate and true posterior distributions at some expense of the mean prediction fitting the data less accurately. Comparing Fig.~\ref{fig:3_county_pp} to Fig.~\ref{fig:VI_conv}, we can see that the coupled, 33-county inversion introduces a bias in the parameter estimates for Bernalillo. This is due to the multitude of NM counties that are sparsely populated and display significant noise in their daily case counts. Despite this effect, the outbreak detection performs well on the 33-county inversion data suggesting that the predictive uncertainties provided by the calibration remain informative.

\emph{To summarize}, reparametrization provided us with the scalability needed to solve the high-dimensional inverse problem conditioned on data from all 33 NM counties.

\section{Discussion}
\label{sec:discussion}

The results in \S~\ref{sec:res} show that MFVI can estimate an infection-rate field and the posterior distribution can be used to produce PPT runs. The MFVI parameter estimates do not quite agree with the AMCMC estimates from the Part I paper\cite{24sr3a,safta2024detecting}, but their effect on the PPT runs is muted, as seen in the comparison in  Fig.~\ref{fig:3_county_pp_vimcmc} and the CRPS summaries in Table~\ref{tab:crps}. Given that the inversion is a smoothing operation, i.e., we learn the infection-rate from historical data, any forecast produced with the estimated infection-rate will be predictive only if the epidemiological dynamics do not change. Consequently, if the forecast and data disagree, it could indicate the arrival of a new wave of infection.

\begin{figure}
    \centering
    \includegraphics[width = 1.0\textwidth]{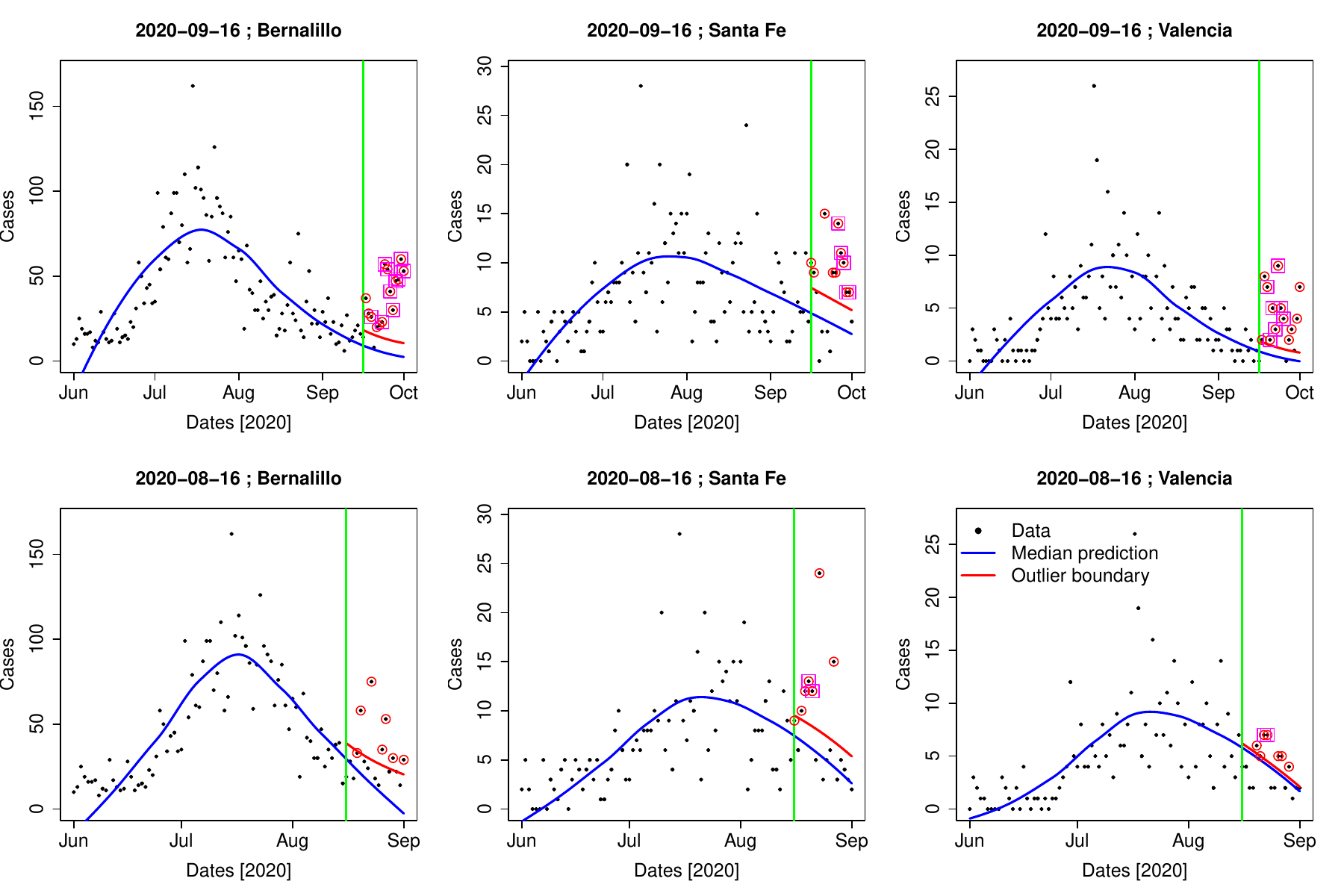}
    \caption{The infection-rate detector implemented using the infection-rate from the joint MFVI estimation of three counties (corresponding to Fig.~\ref{fig:3_county_pp_vimcmc} (top)). The symbols are the case-counts, the solid blue line the median forecast with the infection-rate, post-calibration and the solid red line is the alarm boundary. Outliers are circled and an alarm, corresponding to three consecutive outliers, is encased in a square. The vertical line is the point beyond which we forecast and thus test for the arrival of the Fall 2020 wave. Top row: Testing for arrival after September $15^{\text{th}}$, 2020. Bottom row: Testing for arrival after August $15^{\text{th}}$, 2020.}
    \label{fig:tempDetect03}
\end{figure}

{\bf Temporal detection:} The argument above was used to fashion an outbreak detector using PPT runs in the Part I paper. The detector works as follows. We sample the posterior in the same manner as for PPT runs, and use $\Ypred$ to compute an ``outlier boundary''; in line with our Part I paper, we define it as the ${\rm 99^{th}}$ percentile of the forecasts. $\Ypred$ is computed using data from June $1^{\text{st}}$ to an end date (usually August $15^{\text{th}}$ or September $15^{\text{th}}$) where we test for a change in the epidemiological dynamics. The actual test consists of comparing a two-week-ahead forecast with the data that was observed during that period. Any day with case-counts above the ``outlier boundary'' is deemed an outlier. Three consecutive outlier days cause an ``alarm'', corresponding to an anomalous change in the disease dynamics. Using this detector, in the Part I paper, we found that we could detect the arrival of the Fall 2020 wave correctly, when tested using data up to September $15^{\text{th}}$. When using data up to August $15^{\text{th}}$, a month \emph{before} the arrival of the Fall 2020 wave, we did not encounter a false positive. The method was also compared against a conventional detector\cite{Hohle:2008}, which performed poorly - the reason for this performance is discussed in the Part I paper. The infection-rate estimation was performed using AMCMC, for Bernalillo, Santa Fe and Valencia jointly. In Fig.~\ref{fig:tempDetect03} we repeat the same test, but the infection-rate is estimated using MFVI. The top row depicts the outbreak detector being applied beyond September $15^{\text{th}}$, 2020. We see outliers and alarms for all three counties with a week of September $15^{\text{th}}$ i.e., the Fall 2020 wave was easily detected within a week's worth of data. The bottom row repeats the test, but applied to August $15^{\text{th}}$. We see that Santa Fe and Valencia incur false positives, whereas Bernalillo does not. This implies that the approximations in MFVI may lead to erroneous detections for borderline cases (i.e., counties with small-count data and high variance noise) but areal units with large case-counts might be unaffected.

The reason for the false positives is simple - the data is very noisy and the outbreak detector makes no attempt to reduce the variance in the noise. Comparing the correct detection on the top row of Fig.~\ref{fig:tempDetect03} with the false positives in the bottom row, we see that outliers and alarms are plentiful after September $15^{\text{th}}$ and sporadic after August $15^{\text{th}}$. A slightly more sophisticated detector that performed temporal averaging could eliminate such false positives. We will address this below, using a simple spatio-temporal method.

\begin{figure}
    \centering
    \includegraphics[height = 0.75\textheight]{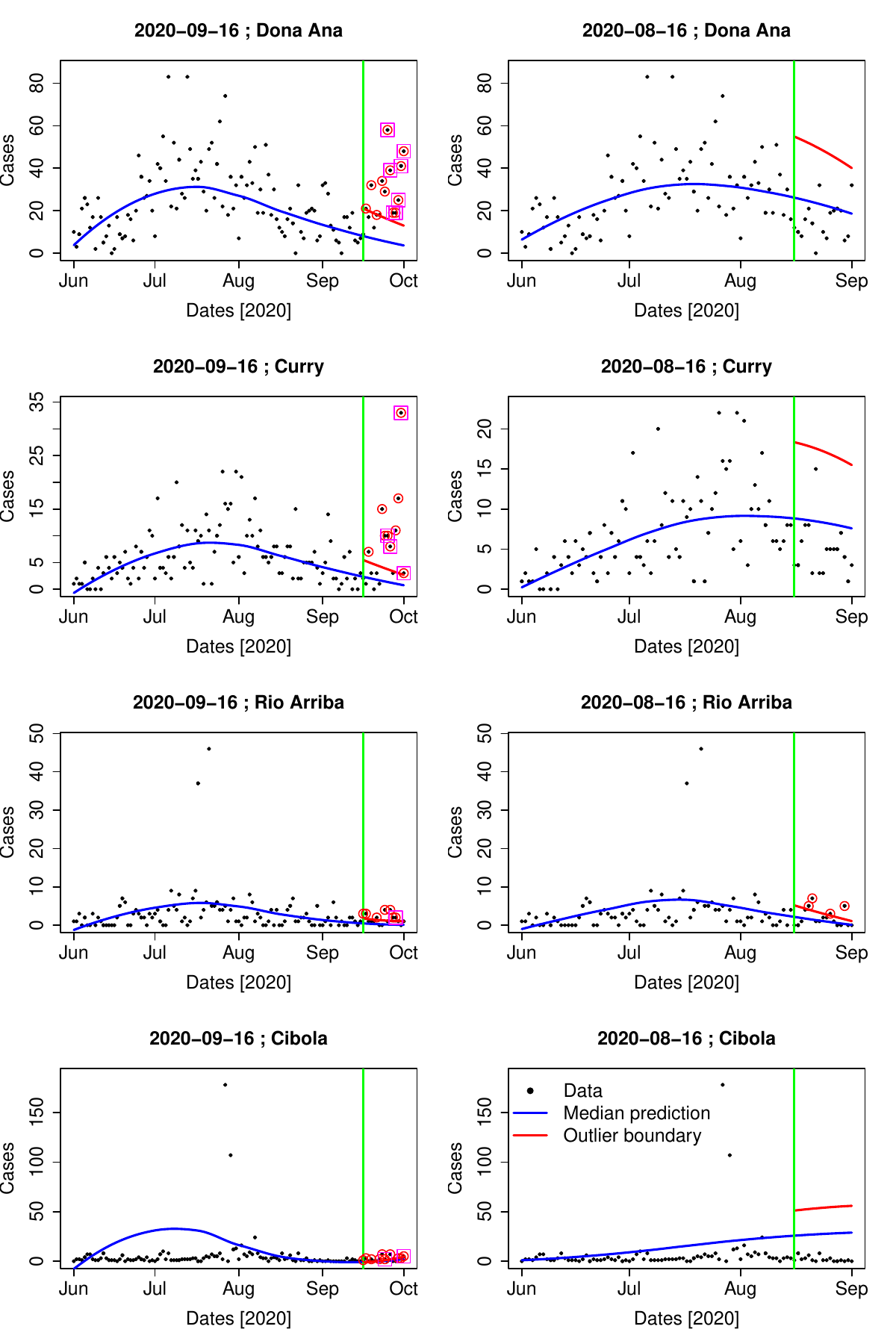}
    \caption{Performance of the outbreak detector for a ``good'' (Do\~{n}a Ana), two ``middling'' (Curry and Rio Arriba) and one ``bad'' county (Cibola), as determined using CRPS in Fig.~\ref{fig:acc}. Left column: Testing for Fall 2020 wave arrival on September $15^{\text{th}}$, 2020. Right column: The same, but detection performed on August $15^{\text{th}}$. We do not see any false positives in the right column.}
    \label{fig:tempDetect33}
\end{figure}
In Fig.~\ref{fig:tempDetect33}, we demonstrate the outbreak detector on a ``good'' (Do\~{n}a Ana), two ``middling'' (Curry and Rio Arriba) and one ``bad'' county (Cibola), as determined using CRPS in Fig.~\ref{fig:acc}. The plotting conventions are the same as in Fig.~\ref{fig:tempDetect03}. We see that we correctly detect the Fall 2020 wave when testing on September $15^{\text{th}}$ and do not incur any false positive on August $15^{\text{th}}$. This performance is rather fortuitous for Rio Arriba and Cibola, which have unexplained case-count peaks in mid-July (Rio Arriba) and in late July - early August (Cibola). It would be difficult to estimate an infection-rate profile for either of these counties independently (this is especially true for Cibola, where the \emph{reported} case-counts show no wave-like structure); however, our spatial GMRF model regularizes the inversion and constructs an infection-rate stably. Forecasts using this infection-rate do not match the case-count data at all, as is clear in the plots for Cibola, but provides us with stable PPT results and a perhaps meaningless detection. However, numerical and algorithmic stability in the face of poor quality data encourage us to believe that the MFVI algorithm and outbreak detection can be automated.

\begin{figure}
    \centerline{
    \includegraphics[width = 0.5\textwidth]{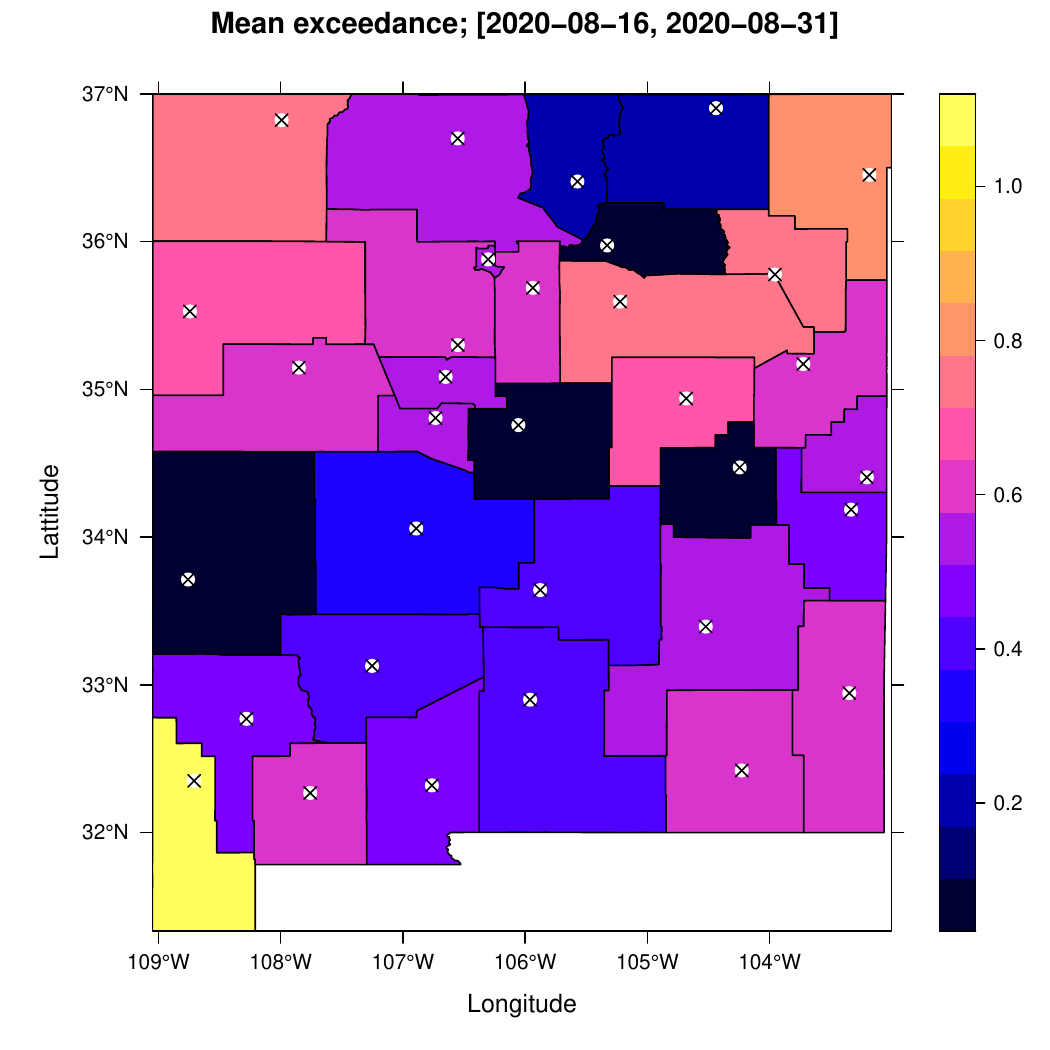}
    \includegraphics[width = 0.5\textwidth]{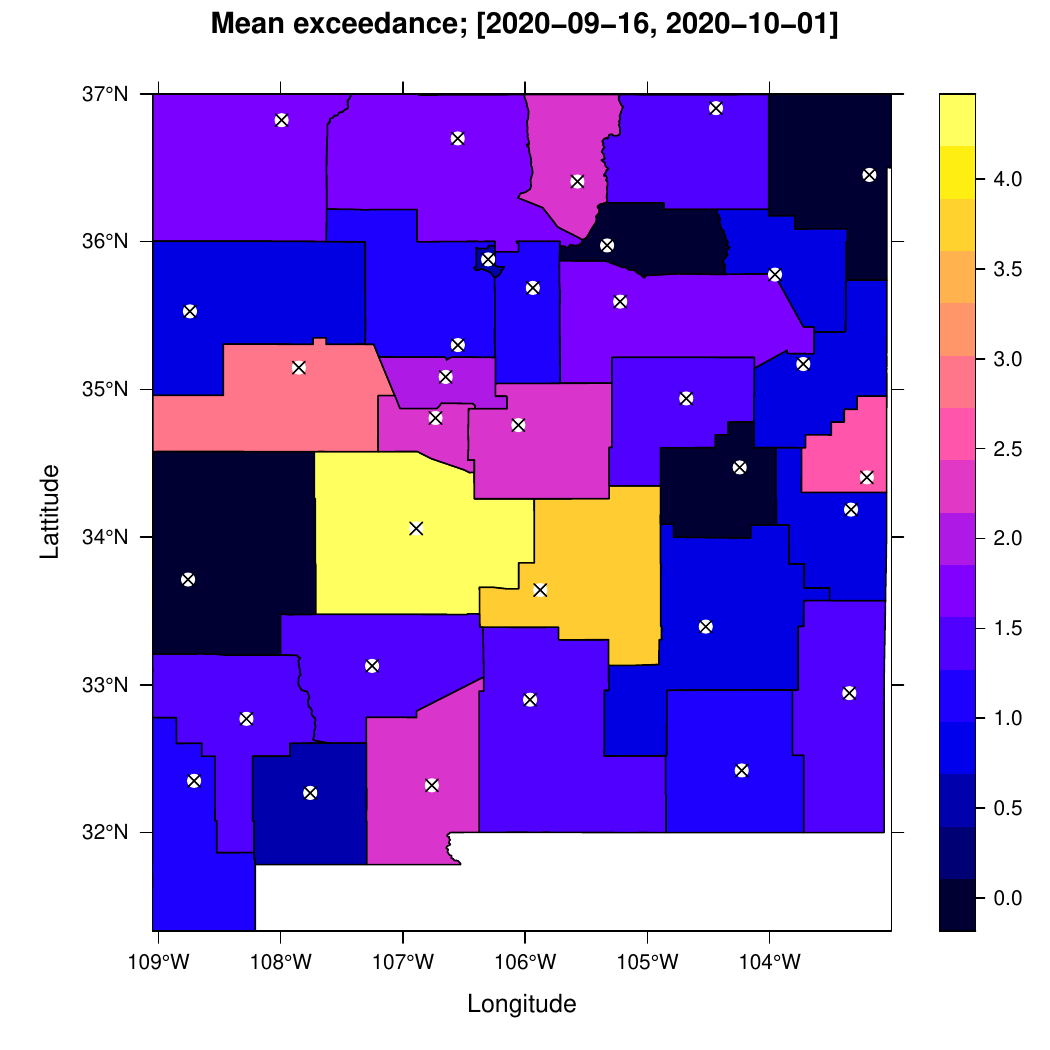} }
    \centerline{
    \includegraphics[width = 0.5\textwidth]{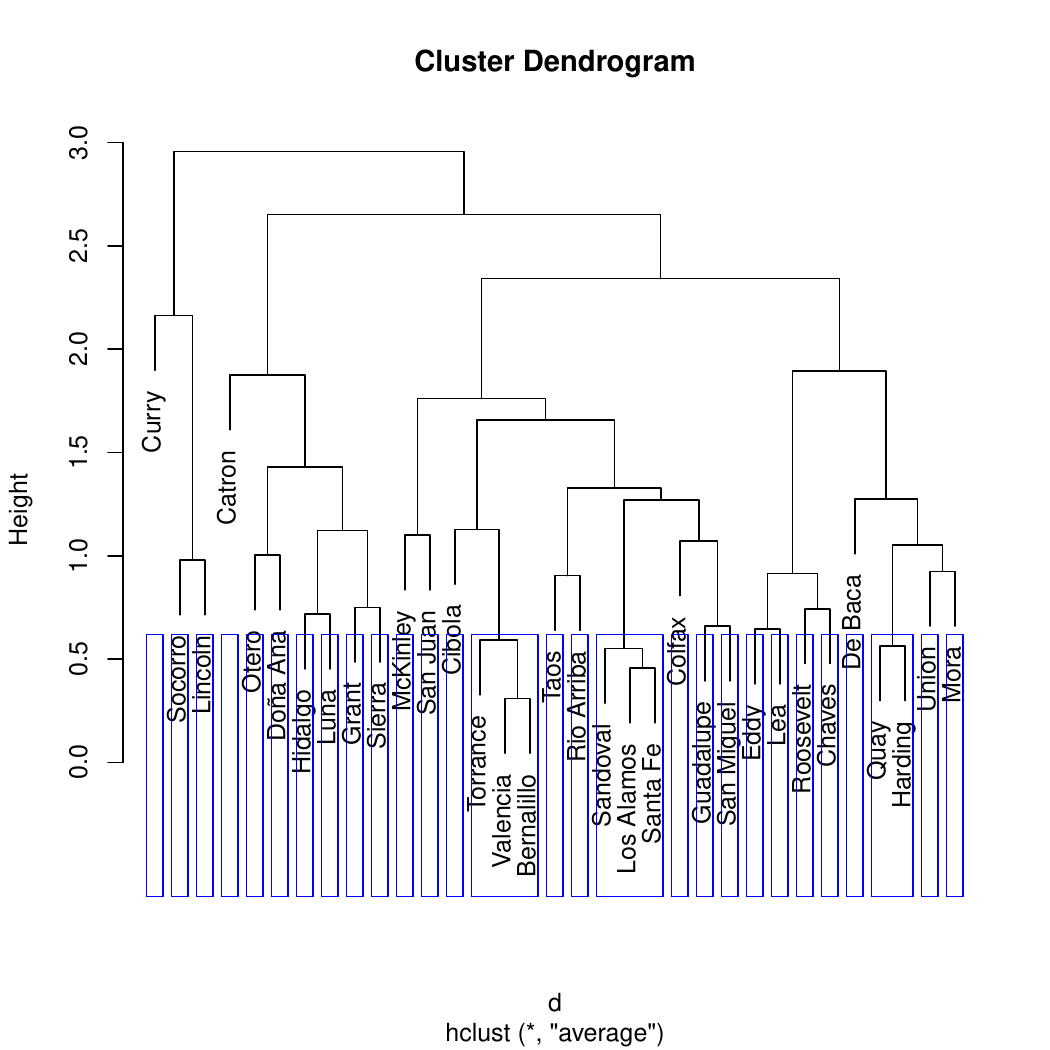}
    \includegraphics[width = 0.5\textwidth]{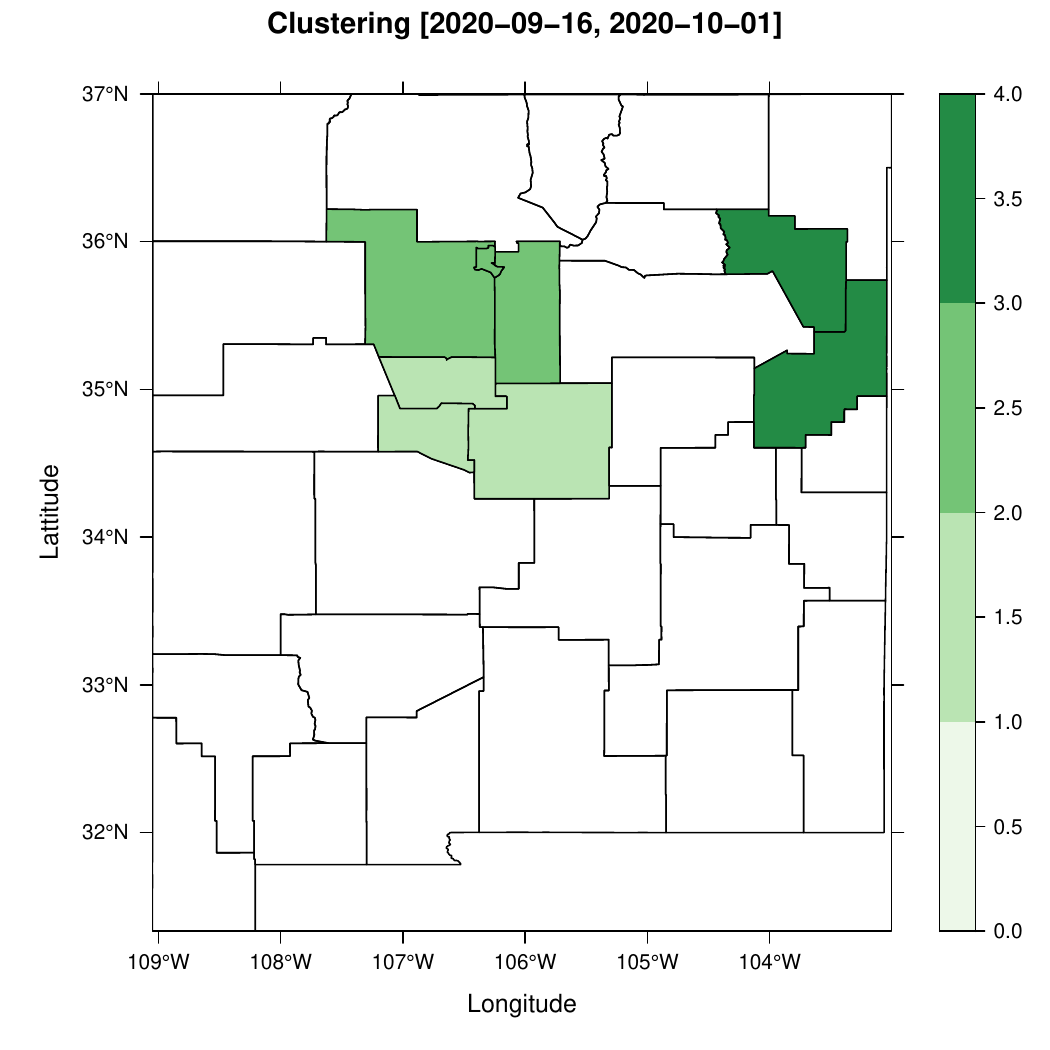}}
    \caption{Top: Plot of mean exceedance $\meanex$ for NM, where $\meanex$ is averaged over August $15^{\text{th}}$ and August $31^{\text{st}}$ (left) before the arrival of the Fall 2020 wave and September $16^{\text{th}}$ and October $1^{\text{st}}$ (right), when the Fall 2020 wave had arrived. Bottom left: Dendrogram from hierarchical clustering, with the cut at a height of 0.6, resulting in clustering of counties. Bottom right: Disease clusters from the dendrogram.}
    \label{fig:clusters}
\end{figure}
{\bf Detecting spatial patterns: } The results above show that the MFVI estimates of the infection-rate have the ability to detect the Fall 2020 wave, especially for the counties in the first quartile (plotted in Fig.~\ref{fig:acc}), though our crude detector, which does not smooth / de-noise the data, might suffer from false positives due to high variance noise in observed case-count. However, the MFVI infection-rate \emph{field} allows prediction in space and time, which allows spatio-temporal assimilation of data. We address this next.

Define ``exceedance'' $\gamma_{r,i} = y^{obs}_{r,i} / y^{(99)}_{r,i}$ for region $r$ and day $i$. Here $y^{obs}_{r,i}$ is the observed case-counts and $y^{(99)}_{r,i}$ is the alarm boundary (the ${\rm 99^{th}}$ percentile computed from $\Ypred$). $\gamma_{r,i} > 1$ would denote an outlier. Let
\[
\meanex = \frac{1}{N_{smooth}} \sum_{i = 1}^{N_{smooth}} \gamma_{r,i}
\]
where $N_{smooth}$ is a time-period over which we will average out the temporal noise and detect spatial patterns. Fig.~\ref{fig:clusters} (top row) shows a map of $\meanex$ computed over a two-week period after August ${\rm 15^{th}}$ (on the left) and a two-week period after September ${\rm 15^{th}}$ (on the right) when the Fall 2020 wave had arrived. We see from the colormap that the $\meanex \leq 1$ before the arrival in all counties except one; specifically, the false positives seen in Fig.~\ref{fig:tempDetect03} for Santa Fe and Valencia are no longer visible. After the arrival, we see clear spatial structures on the right. Thus simply averaging the data in time provides a more robust detection, though with a loss of timeliness. However, we see two adjacent counties (in shades of yellow) with very high $\meanex$.

Epidemiological activity spreads due to population mixing, and as we found in our Part I paper, this is largely between adjoining NM counties (because of their large spatial expanse). Therefore adjacent counties are expected to have similar epidemiological behavior. If the epidemiological activity is represented using a derived / inferred / estimated quantity, and adjoining areal units differ greatly, it is either a consequence of erroneous estimation or an artifact of noise in data i.e., spatial autocorrelation might be a simple mechanism for removing erratic behavior in data. To this end, we subject the $\meanex$ map in Fig.~\ref{fig:clusters} (top right) to clustering. The centroids of the counties serve as the spatial feature whereas $\meanex$ serves as a measure of how strong the Fall 2020 wave is, in each county. The data was Z-scored and subjected to hierarchical clustering using the R Statistical Software\cite{Manual:R} (R version 4.3.2 (2023-10-31)) package \texttt{stats}, specifically the function \texttt{hclust()}. The resulting dendrogram is in Fig.~\ref{fig:clusters} (bottom left), which is cut at a height of 0.6 (corresponding to a quantile of 0.15) to reveal 3 clusters of counties. These are plotted in Fig.~\ref{fig:clusters} (bottom right) and reveal the same clustering as was seen in Fig.~\ref{fig:clusters} (top right). However, the counties in yellow were eliminated - their level of $\meanex$, though high, were too different and thus violated the need for spatial autocorrelation. Two of the clusters surround the county of Bernalillo, the population center of NM, where COVID-19 was very active in the city of Albuquerque. 

\section{Conclusion}
\label{sec:concl}
In this paper, we have developed a scalable but approximate method to estimate the infection-rate field of an outbreak,  defined over a collection of areal units. The method was demonstrated using COVID-19 data from the counties of NM. The purpose of estimating the infection-rate field was to detect a sudden change in the epidemiological dynamics, corresponding to the arrival of a new wave of infections. Contemporary methods use case-counts to perform this detection and are plagued by stochasticity and reporting errors in the data, especially when case-counts are low (as may be expected from sparsely-populated regions); in contrast, the infection-rate is governed by human mixing patterns that do not vary erratically day-to-day. Our method is based on mean-field variational inference (MFVI), but required some innovations for computational efficiency and for enforcing non-negativity constraints in the variables being estimated. The MFVI builds on a disease modeling formalism that was developed in the Part I paper\cite{24sr3a,safta2024detecting}, which also addressed the issue of spatial autocorrelation in the COVID-19 dataset with a Gaussian Markov Random field. This previous paper used Adaptive Markov chain Monte Carlo (AMCMC) as the estimation algorithm and could not scale to all the counties of NM.

The MFVI method obtains its scalability (in terms of the number of variables being estimated) by imputing a parametrized form for the posterior density, a set of independent Gaussians in our case; thereafter we estimate their means and standard deviations using scalable gradient-based algorithms. We compared the predictive skill of disease models that used the infection-rate field estimated using AMCMC versus MFVI to answer the research questions posed in \S~\ref{sec:intro}. Our findings are listed below.
\begin{itemize}
    \item We find that the predictive skills of AMCMC- and MFVI-calibrated epidemiological models are similar, even though the parameters' posterior distributions estimated by the two method are somewhat different. The uncertainties in the parameters estimated from MFVI are too small to be credible, in line with previous findings\cite{Han:2019}. Both the approaches also estimate the noise in the data (i.e., the component of data variability that cannot be represented by the disease model), and the MFVI estimate is far larger than the AMCMC counterpart. In this manner MFVI compensates for its spuriously low uncertainty estimates in the parameters of the disease model, and achieves similar levels of predictive skill as the AMCMC method (which makes no approximations). 
    \item The GMRF spatial model, developed in the Part I paper\cite{24sr3a,safta2024detecting}, plays an important role in stabilizing the inversion. It regularizes the estimation problem with a Gaussian Markov Random Field, and is sufficient to stabilize the estimation when the observed case-counts from certain areal units (counties) bear no resemblance to the waxing and waning of the COVID-19 pandemic that is clearly observed in case-count data aggregated to the state level.  Counties like Cibola and Rio Arriba, where errors  / shortcomings in the data do not allow a reliable estimation of the infection-rate are regularised by it, and in the case of Rio Arriba also provides a credible forecast of the epidemiological dynamics. This robustness to occasional low quality data, which is to be expected for large inversions, allows the automation of the estimation process without any manual ``cleaning'' of the data and imputations of ``cleaned'' data values.
    \item The infection-rate field, estimated using MFVI, is used to detect the arrival of the Fall 2020 wave of COVID-19 in NM. To do so, we design a crude temporal anomaly detector which contrasts model forecasts with observed data; large discrepancies imply a change in epidemiological dynamics from the past. We had no difficulty in detecting the wave when it was present, but suffered from false positives when tests were conducted using data from before the arrival. These false positives were caused by high-variance noise in low case-count data and could be removed by temporal averaging. However, by doing so, our detections were no longer very timely. In addition, the availability of an infection-rate \emph{field} allowed us to exploit the spatial autocorrelation to remove counties with spuriously high levels of epidemiological activity when their neighbors were quiescent. It is clear that MFVI yields useful information about the outbreak signature and does so in a scalable fashion. 
\end{itemize}

Note that we made no attempt to design a proper anomaly detector with tunable parameters to trade-off specificity versus sensitivity and plot Receiver Operation Characteristic curves; our aim was to merely test of the existence of certain information in the infection-rate estimates. However, the study revealed that a proper anomaly detector would either have to smooth out high-variance noise in low case-count data, or be formulated using Negative Binomial or Poisson assumptions; these will be necessary to suppress false positives when the data has high-variance noise. False negatives, on the other hand, may not require any special considerations; this is certainly true for the NM COVID-19 dataset.

\clearpage
\section*{\bf Author contributions}
Wyatt Bridgman formulated the problem, wrote the software to solve it, generated the figures and wrote the paper. Cosmin Safta assisted with variational inference, interpretation of results, and contributed to writing the paper. Jaideep Ray posed the problem, assisted with the epidemiological interpretation, and suggested the calibration approach and metrics.

\section*{Acknowledgments}
This paper (SAND2024-08888O) describes objective technical results and analysis. Any subjective views or opinions that might be expressed in the paper do not necessarily represent the views of the U.S. Department of Energy or the United States Government. This article has been authored by an employee of National Technology \& Engineering Solutions of Sandia, LLC under Contract No. DE-NA0003525 with the U.S. Department of Energy (DOE). The employee owns all right, title and interest in and to the article and is solely responsible for its contents. The United States Government retains and the publisher, by accepting the article for publication, acknowledges that the United States Government retains a non-exclusive, paid-up, irrevocable, world-wide license to publish or reproduce the published form of this article or allow others to do so, for United States Government purposes. The DOE will provide public access to these results of federally sponsored research in accordance with the DOE Public Access Plan https://www.energy.gov/downloads/doe-public-access-plan.  

\section*{\bf Financial disclosure}
None reported.

\section*{\bf Conflict of interest}
The authors declare no potential conflict of interests.


\appendix
\label{sec:appendix}
\section{Variational Inference}
\label{sec:appendix-VI}

\subsection{Score gradients of the ELBO}
\label{sec:appendix-ELBO-score}

We briefly review the score estimator, or black-box, approach to estimating the gradient of the ELBO. Recall that the ELBO is given by \ref{eq:ELBO} which, for the sake of clarity, can be written in a more generic form
\begin{equation}
    \mathcal{L}(\boldsymbol{\phi}) = \mathbb{E}_{q(\boldsymbol{\theta};\boldsymbol{\phi})} \left[ f(\boldsymbol{\theta}) \right] 
\end{equation}
where $f(\boldsymbol{\theta})$ encapsulates the dependence on the random vector $\boldsymbol{\theta}$. To derive an estimator for the gradient, we can carry out the following manipulations
\begin{align}
    \nabla_{\boldsymbol{\phi}} \mathcal{L}(\boldsymbol{\phi}) &= \nabla_{\boldsymbol{\phi}} \mathbb{E}_{q(\boldsymbol{\theta};\boldsymbol{\phi})} \left[ f(\boldsymbol{\theta}) \right] \\
    &= \nabla_{\boldsymbol{\phi}} \int q(\boldsymbol{\theta};\boldsymbol{\phi}) f(\boldsymbol{\theta})  d\,\boldsymbol{\theta} \\
    &=  \int \nabla_{\boldsymbol{\phi}} q(\boldsymbol{\theta};\boldsymbol{\phi}) f(\boldsymbol{\theta})  d\,\boldsymbol{\theta} \\
    &= \int q(\boldsymbol{\theta};\boldsymbol{\phi}) \frac{\nabla_{\boldsymbol{\phi}} q(\boldsymbol{\theta};\boldsymbol{\phi})}{q(\boldsymbol{\theta};\boldsymbol{\phi})}f(\boldsymbol{\theta}) d\,\boldsymbol{\theta} \\
    &= \mathbb{E}_{q(\boldsymbol{\theta};\boldsymbol{\phi})} \left[ f(\boldsymbol{\theta}) \nabla_{\boldsymbol{\phi}} \log q(\boldsymbol{\theta};\boldsymbol{\phi}) \right]
\end{align}
Hence, the gradient can be expressed as an expectation with respect to $q(\boldsymbol{\theta};\boldsymbol{\phi})$ where only the log of the surrogate posterior needs to be differentiated with respect to the variational parameters $\boldsymbol{\phi}$.

\subsection{Reparametrization gradients of the ELBO}
\label{sec:appendix-ELBO-reparam}

The likelihood and log likelihood are given by
\begin{align}
    p(\mathcal{D} \vert \boldsymbol{\theta}) &= \prod_{i=1}^{N_d} 2\pi^{-N_r / 2} \det(\mathbf{\Sigma}_i)^{-1/2} \exp \left( -\frac{1}{2} (\mathbf{y}_i^{(o)} - \mathbf{y}_i)^T \mathbf{\Sigma}_i^{-1} (\mathbf{y}_i^{(o)} - \mathbf{y}_i) \right) \\
    l(\boldsymbol{\theta}) &= -\frac{N_d N_r 2 \pi}{2} -\frac{1}{2} \sum_{i=1}^{N_d} \log \det(\mathbf{\Sigma}_i) +(\mathbf{y}_i^{(o)} - \mathbf{y}_i)^T \mathbf{\Sigma}_i^{-1} (\mathbf{y}_i^{(o)} - \mathbf{y}_i)
\end{align}
Using the reparametrization trick, we can write the ELBO \eqref{eq:ELBO} and its gradient in the form
\begin{align}
    \mathcal{L}(\boldsymbol{\phi}) &= -\mathbb{H}[q(\boldsymbol{\theta};\boldsymbol{\phi})] - \mathbb{E}_{q(\boldsymbol{\epsilon}) } [ \log p(\mathcal{D} \vert \boldsymbol{\theta}(\boldsymbol{\epsilon},\boldsymbol{\phi})) + \log p(\boldsymbol{\theta}(\boldsymbol{\epsilon},\boldsymbol{\phi}))] \\
    \nabla_{\boldsymbol{\phi}} \mathcal{L}(\boldsymbol{\phi}) &= - \nabla_{\boldsymbol{\phi}} \mathbb{H}[q(\boldsymbol{\theta};\boldsymbol{\phi})] - \mathbb{E}_{q(\boldsymbol{\epsilon}) } [\nabla_{\boldsymbol{\phi}}  \log p(\mathcal{D} \vert \boldsymbol{\theta}(\boldsymbol{\epsilon},\boldsymbol{\phi})) + \nabla_{\boldsymbol{\phi}} \log p(\boldsymbol{\theta}(\boldsymbol{\epsilon},\boldsymbol{\phi}))]
\end{align}
where $\boldsymbol{\phi} = (\boldsymbol{\mu},\boldsymbol{\rho})$, $\boldsymbol{\theta} = \boldsymbol{\mu} + \boldsymbol{\sigma}(\boldsymbol{\rho}) \odot \boldsymbol{\epsilon}$ with $\boldsymbol{\epsilon} \sim \mathcal{N}(\mathbf{0},\mathbf{I})$. Here, $\sigma$ is a positive transformation of the unconstrained variable $\boldsymbol{\rho}$ to ensure the variance is constrained to be positive. A Monte Carlo estimator of the gradient can then be written as
\begin{align}
    \nabla_{\boldsymbol{\phi}} \mathcal{L}(\boldsymbol{\phi}) &\approx - \nabla_{\boldsymbol{\phi}} \mathbb{H}[q(\boldsymbol{\theta};\boldsymbol{\phi})]  -\frac{1}{N_s}\sum_{i=1}^{N_s} \nabla_{\boldsymbol{\phi}}  \log p(\mathcal{D} \vert \boldsymbol{\theta}(\boldsymbol{\epsilon}_i,\boldsymbol{\phi})) + \nabla_{\boldsymbol{\phi}} \log p(\boldsymbol{\theta}(\boldsymbol{\epsilon}_i,\boldsymbol{\phi})) \\
    &= - \nabla_{\boldsymbol{\phi}} \mathbb{H}[q(\boldsymbol{\theta};\boldsymbol{\phi})]  -\frac{1}{N_s}\sum_{i=1}^{N_s} \left( \nabla_{\boldsymbol{\theta}}  \log p(\mathcal{D} \vert \boldsymbol{\theta}(\boldsymbol{\epsilon}_i,\boldsymbol{\phi}))  + \nabla_{\boldsymbol{\theta}} \log p(\boldsymbol{\theta}(\boldsymbol{\epsilon}_i,\boldsymbol{\phi})) \right) \odot \nabla_{\boldsymbol{\phi}} \boldsymbol{\theta}(\boldsymbol{\epsilon}_i,\boldsymbol{\phi})
\end{align}
where the last line is given by the chain rule and the fact that $\boldsymbol{\theta}$ is defined by an element-wise transformation of $\boldsymbol{\phi}$. Observe that
\begin{align}
    \nabla_{\boldsymbol{\mu}}  \boldsymbol{\theta} &= \mathbf{1} \\
    \nabla_{\boldsymbol{\rho}} \boldsymbol{\theta} &= \nabla_{\boldsymbol{\rho}} \boldsymbol{\sigma}(\boldsymbol{\rho}) \odot \boldsymbol{\epsilon}_i
    \\
    \nabla_{\boldsymbol{\mu}} \mathbb{H}[q(\boldsymbol{\theta};\boldsymbol{\phi})] &= \mathbf{0} \\
    \nabla_{\boldsymbol{\rho}} \mathbb{H}[q(\boldsymbol{\theta};\boldsymbol{\phi})] &= \frac{1}{\boldsymbol{\sigma}(\boldsymbol{\rho})} \odot \nabla_{\boldsymbol{\rho}} \boldsymbol{\sigma}(\boldsymbol{\rho})
\end{align}
so that it remains to compute the gradients of the log-likelihood and prior. 

\subsection{Gradients of the Log Likelihood}
\label{sec:appendix-grad-loglik}

As the log likelihood factors independently across the data $i=1,\ldots,N_d$, it suffices to compute the gradients $\nabla_{\boldsymbol{\theta}} \log \det \mathbf{\Sigma}_i$ and $\nabla_{\boldsymbol{\theta}} (\mathbf{y}_i^{(o)} - \mathbf{y}_i)^T \mathbf{\Sigma}_i^{-1} (\mathbf{y}_i^{(o)} - \mathbf{y}_i)$ for a particular day $i$. The differentials of the these two terms are computed using matrix calculus \cite{Petersen:2008} as
\begin{align}
    \partial \log \det \mathbf{\Sigma}_i &= \text{Tr}(\mathbf{\Sigma}_i^{-1} \partial \mathbf{\Sigma}_i) \\
    \partial (\mathbf{y}_i^{(o)} - \mathbf{y}_i)^T \mathbf{\Sigma}_i^{-1} (\mathbf{y}_i^{(o)} - \mathbf{y}_i) &= -(\mathbf{y}_i^{(o)} - \mathbf{y}_i)^T \mathbf{\Sigma}_i^{-1} (\partial \mathbf{\Sigma}_i )\mathbf{\Sigma}_i^{-1}(\mathbf{y}_i^{(o)} - \mathbf{y}_i) - 2(\mathbf{y}_i^{(o)} - \mathbf{y}_i)^T \mathbf{\Sigma}_i^{-1} \partial \mathbf{y}_i 
\end{align}
where the differential of $\mathbf{\Sigma}_i$ is
\begin{align}
    \partial \mathbf{\Sigma}_i &= (\partial \tau_{\Phi}) [\mathbf{D} - \lambda_{\Phi} \mathbf{W}]^{-1} - (\partial \lambda_{\Phi}) \tau_{\Phi} [\mathbf{D} - \lambda_{\Phi} \mathbf{W}]^{-1} \mathbf{W} [\mathbf{D} - \lambda_{\Phi} \mathbf{W}]^{-1} \\
    &+ 2 \text{diag}(\sigma_a + \sigma_m \mathbf{y}_i)[(\partial \sigma_a) \mathbf{I} + (\partial \sigma_m) \text{diag}(\mathbf{y}_i) + \sigma_m \text{diag}(\partial \mathbf{y}_i) ]
\end{align} 
 We have the following derivatives
\begin{itemize}
    \item \textbf{Case:} $\hat{\mathbf{m}}_r^T = (\hat{t}_0^r, \hat{N}^r, \hat{k}^r, \hat{\theta}^r)$ are the unconstrained model variables for region $r$ and $\hat{\theta}_r$ is a particular variable.
    \begin{align}
        \partial_{\hat{\theta}_r} \mathbf{y}_i &= (0,\ldots,\partial_{\theta_r} y_r(i;t_0^r, N_r, k^r,\theta^r) \theta'_i(\hat{\theta}_i),\ldots,0) \\
        \nabla_{\hat{\mathbf{m}}_r} \log \det \mathbf{\Sigma}_i &= 2 \sigma_m [\mathbf{\Sigma}_i^{-1}]_{i i} \nabla_{\hat{\mathbf{m}}_r} y_r(i;t_0^r, N^r, k^r,\theta^r)  \\
        \partial_{\hat{\theta}_r} (\mathbf{y}_i^{(o)} - \mathbf{y}_i)^T \mathbf{\Sigma}_i^{-1} (\mathbf{y}_i^{(o)} - \mathbf{y}_i) &= - 2 \sigma_m(\mathbf{y}_i^{(o)} - \mathbf{y}_i)^T \mathbf{\Sigma}_i^{-1}  \text{diag}(\sigma_a + \sigma_m \mathbf{y}_i) \text{diag}(\partial_{\hat{\theta}_r} \mathbf{y}_i) \mathbf{\Sigma}_i^{-1} (\mathbf{y}_i^{(o)} - \mathbf{y}_i) \\
        &- 2(\mathbf{y}_i^{(o)} - \mathbf{y}_i)^T \mathbf{\Sigma}_i^{-1} \partial_{\hat{\theta}_r} \mathbf{y}_i
    \end{align}

    \item \textbf{Case:}  $\hat{\theta}_r = \hat{\tau}_{\Phi}$
    \begin{align}
        \partial_{\hat{\theta}_r} \mathbf{y}_i &= \mathbf{0} \\
        \partial_{\hat{\theta}_r} \log \det \mathbf{\Sigma}_i &= \tau'_{\Phi}(\hat{\tau}_{\Phi}) \text{Tr}(\mathbf{\Sigma}_i^{-1} [\mathbf{I}-\lambda_{\Phi} \mathbf{W}]^{-1}) \\
        \partial_{\hat{\theta}_r} (\mathbf{y}_i^{(o)} - \mathbf{y}_i)^T \mathbf{\Sigma}_i^{-1} (\mathbf{y}_i^{(o)} - \mathbf{y}_i) &= \tau'_{\Phi}(\hat{\tau}_{\Phi}) (\mathbf{y}_i^{(o)} - \mathbf{y}_i)^T \mathbf{\Sigma}_i^{-1} [\mathbf{I}-\lambda_{\Phi} \mathbf{W}]^{-1} \mathbf{\Sigma}_i^{-1} (\mathbf{y}_i^{(o)} - \mathbf{y}_i)
    \end{align}
    \item \textbf{Case:} $\hat{\theta}_r = \hat{\lambda}_{\Phi}$
    \begin{align}
        \partial_{\hat{\theta}_r} \mathbf{y}_i &= \mathbf{0} \\
        \partial_{\hat{\theta}_r} \log \det \mathbf{\Sigma}_i &= \lambda'_{\Phi}(\hat{\lambda}_{\Phi}) \tau_{\Phi} \text{Tr} (\mathbf{\Sigma}_i^{-1} [\mathbf{I}-\lambda_{\Phi} \mathbf{W}]^{-1} \mathbf{W} [\mathbf{I}-\lambda_{\Phi} \mathbf{W}]^{-1}) \\
        \partial_{\hat{\theta}_r} (\mathbf{y}_i^{(o)} - \mathbf{y}_i)^T \mathbf{\Sigma}_i^{-1} (\mathbf{y}_i^{(o)} - \mathbf{y}_i) &= \lambda'_{\Phi}(\hat{\lambda}_{\Phi}) \tau_{\Phi} (\mathbf{y}_i^{(o)} - \mathbf{y}_i)^T \mathbf{\Sigma}_i^{-1} [\mathbf{I}-\lambda_{\Phi} \mathbf{W}]^{-1} \mathbf{W}[\mathbf{I}-\lambda_{\Phi} \mathbf{W}]^{-1} \mathbf{\Sigma}_i^{-1} (\mathbf{y}_i^{(o)} - \mathbf{y}_i)
    \end{align}
    \item \textbf{Case:} $\hat{\theta}_r = \hat{\sigma}_a$
    \begin{align}
        \partial_{\hat{\theta}_r} \mathbf{y}_i &= \mathbf{0} \\
        \partial_{\hat{\theta}_r} \log \det \mathbf{\Sigma}_i &= 2 \sigma'_a(\hat{\sigma}_a) \text{Tr}(\mathbf{\Sigma}_i^{-1} \text{diag}(\sigma_a + \sigma_m \mathbf{y}_i)) \\
        \partial_{\hat{\theta}_r} (\mathbf{y}_i^{(o)} - \mathbf{y}_i)^T \mathbf{\Sigma}_i^{-1} (\mathbf{y}_i^{(o)} - \mathbf{y}_i) &= -2 \sigma'_a(\hat{\sigma}_a) (\mathbf{y}_i^{(o)} - \mathbf{y}_i)^T \mathbf{\Sigma}_i^{-1} \text{diag}(\sigma_a + \sigma_m \mathbf{y}_i) \mathbf{\Sigma}_i^{-1} (\mathbf{y}_i^{(o)} - \mathbf{y}_i)
    \end{align}
    \item \textbf{Case:}$\hat{\theta}_r = \hat{\sigma}_m$
    \begin{align}
        \partial_{\hat{\theta}_r} \mathbf{y}_i &= \mathbf{0} \\
        \partial_{\hat{\theta}_r} \log \det \mathbf{\Sigma}_i &= 2 \sigma'_m(\hat{\sigma}_m) \text{Tr}(\mathbf{\Sigma}_i^{-1} \text{diag}((\sigma_a + \sigma_m \mathbf{y}_i) \odot \mathbf{y}_i)) \\
        \partial_{\hat{\theta}_r} (\mathbf{y}_i^{(o)} - \mathbf{y}_i)^T \mathbf{\Sigma}_i^{-1} (\mathbf{y}_i^{(o)} - \mathbf{y}_i) &= -2 \sigma'_m(\hat{\sigma}_m) (\mathbf{y}_i^{(o)} - \mathbf{y}_i)^T \mathbf{\Sigma}_i^{-1} \text{diag}((\sigma_a + \sigma_m \mathbf{y}_i) \odot \mathbf{y}_i)) \mathbf{\Sigma}_i^{-1} (\mathbf{y}_i^{(o)} - \mathbf{y}_i)
    \end{align}
\end{itemize}

The variable transformations are listed in Table~\ref{tab:variable-transformations}. 
\begin{table}[h]
\centering
\begin{tabular}{ |c|c|c| }  
    \hline
    \textbf{Variable} & \textbf{Constraint} & \textbf{Transformation} \\ 
    \hline
    $t_0^r$ & None & $id(\hat{t}_0^r)$ \\ 
    $N^r$ & $N^r \geq 0$ & $\exp(\hat{N}^r)$ \\ 
    $k^r$ & $k^r \geq 2$ & $s(\hat{k}^r)+2$ \\
    $\theta^r$ & $k^r \geq \epsilon$ & $s(\hat{\theta}^r)+\epsilon$ \\
    $\tau_{\Phi}$ & $\tau_{\Phi} \geq 0$ & $\exp(\hat{\tau}_{\Phi})$ \\
    $\lambda_{\Phi}$ & $0 \leq \lambda_{\Phi} \leq 1$ & $(1+\exp(-\hat{\lambda}_{\Phi}))^{-1}$ \\
    $\sigma_a$ & $\sigma_a \geq 0$ & $\exp(\hat{\sigma}_a)$ \\
    $\sigma_m$ & $\sigma_m \geq 0$ & $\exp(\hat{\sigma}_m)$ \\
    \hline
\end{tabular}
\caption{Variable transformations. $s(x):\log(\exp(x)+1)$ is the \textit{softplus} function and satisfies $s(x) \in [0,1]$.}
\label{tab:variable-transformations}
\end{table}

\subsection{Approximation of model predictions and gradients via quadrature}
\label{sec:appendix-grad-model}

The model predictions $\mathbf{y}_i$, given by \eqref{eq:model-pred}, involve a convolution integral that cannot be expressed in closed form. Hence, we approximate the predictions by integral quadrature
\begin{align}
y_r(i;t_0^r, N^r, k^r,\theta^r)  &= \int_{t_0^r}^{t_i} N_r f_{inf}(\tau-t_0^r;k^r,\theta^r)
    (F_{inc}(t_i-\tau;\mu,\sigma)-(F_{inc}(t_{i-1}-\tau;\mu,\sigma))d\,\tau \\
    &\approx \sum_{j=1}^n N_r w_j f_{inf}(\tau_j-t_0^r;k^r,\theta^r)
    (F_{inc}(t_i-\tau_j;\mu,\sigma)-(F_{inc}(t_{i-1}-\tau_j;\mu,\sigma))
\end{align}
where $w_j$, $\tau_j$ are quadrature weights and points given by a method such as Gaussian quadrature.
As the function $(F_{inc}(t_i-\tau;\mu,\sigma)-(F_{inc}(t_{i-1}-\tau;\mu,\sigma))$ does not depend on parameters $\boldsymbol{\theta}$, we can write it as $\tilde{F}_{inc}(\tau)$ for simplicity of notation. By the Leibniz integral rule, we can write the derivatives of the model predictions as 
\begin{align}
    \partial_{\hat{\theta}_r} \mathbf{y}_i &= \int_{t_0^r}^{t_i} [\partial_{\hat{\theta}_r} N_r f_{inf}(\tau-t_0^r;k^r,\theta^r)] \tilde{F}_{inc}(\tau)d\,\tau &\quad \text{if $\hat{\theta}_r \neq t_0^r$} \label{eq:leibniz-integral-1}\\ 
    \partial_{\hat{\theta}_r} \mathbf{y}_i &= \int_{t_0^r}^{t_i} [\partial_{\hat{\theta}_r} N_r f_{inf}(\tau-t_0^r;k^r,\theta^r)] \tilde{F}_{inc}(\tau)d\,\tau -f_{inf}(0;k^r,\theta^r) &\quad \text{if $\hat{\theta}_r = t_0^r$}
    \label{eq:leibniz-integral-2}
\end{align}
This requires that the functions $N_r f_{inf}(\tau-t_0^r;k^r,\theta^r)] \tilde{F}_{inc}(\tau)$ and $\partial_{\hat{\theta}_r} N_r f_{inf}(\tau-t_0^r;k^r,\theta^r)] \tilde{F}_{inc}(\tau)$ are continuous in $\tau$ and $\hat{\theta}_r$ in a region of the $\tau$-$\hat{\theta}_r$ plane including $t_0 \leq \tau \leq t_i$, a condition that's easily met by ensuring certain constraints are satisfied by the parameters $(t_0^r,N^r,k^r,\theta^r)$ for $r=1,\ldots,N_r$. These constraints are enforced via the variable transformations in \eqref{eq:mean_field_Gaussian_pf}. Hence, we can approximate \eqref{eq:leibniz-integral-1} and \eqref{eq:leibniz-integral-2} via quadrature in the form
\begin{align}
    \partial_{\hat{\theta}_r} \mathbf{y}_i &\approx \sum_{j=1}^n w_j [\partial_{\hat{\theta}_r} N_r f_{inf}(\tau_j-t_0^r;k^r,\theta^r)] \tilde{F}_{inc}(\tau_j) &\quad \text{if $\hat{\theta}_r \neq t_0^r$}\\ 
    \partial_{\hat{\theta}_r} \mathbf{y}_i &\approx \sum_{j=1}^n w_j [\partial_{\hat{\theta}_r} N_r f_{inf}(\tau_j-t_0^r;k^r,\theta^r)] \tilde{F}_{inc}(\tau_j) - f_{inf}(0;k^r,\theta^r) &\quad \text{if $\hat{\theta}_r = t_0^r$}
\end{align}
Hence, in this implementation of VI, gradients of the ELBO have a pseudo-analytic form where ``outer" gradients of the log likelihood with respect to model predictions are exact and ``inner" gradients of the model predictions with respect to parameters are approximated via quadrature. This allows for accurate gradient approximations that can be caculated efficiently leading to a scalable VI algorithm that can be applied to the high-dimensional inverse problem for the outbreak model.

\section{Infection-rate estimates}

\begin{figure}[!t]
	\centering
    \includegraphics[width=\textwidth]{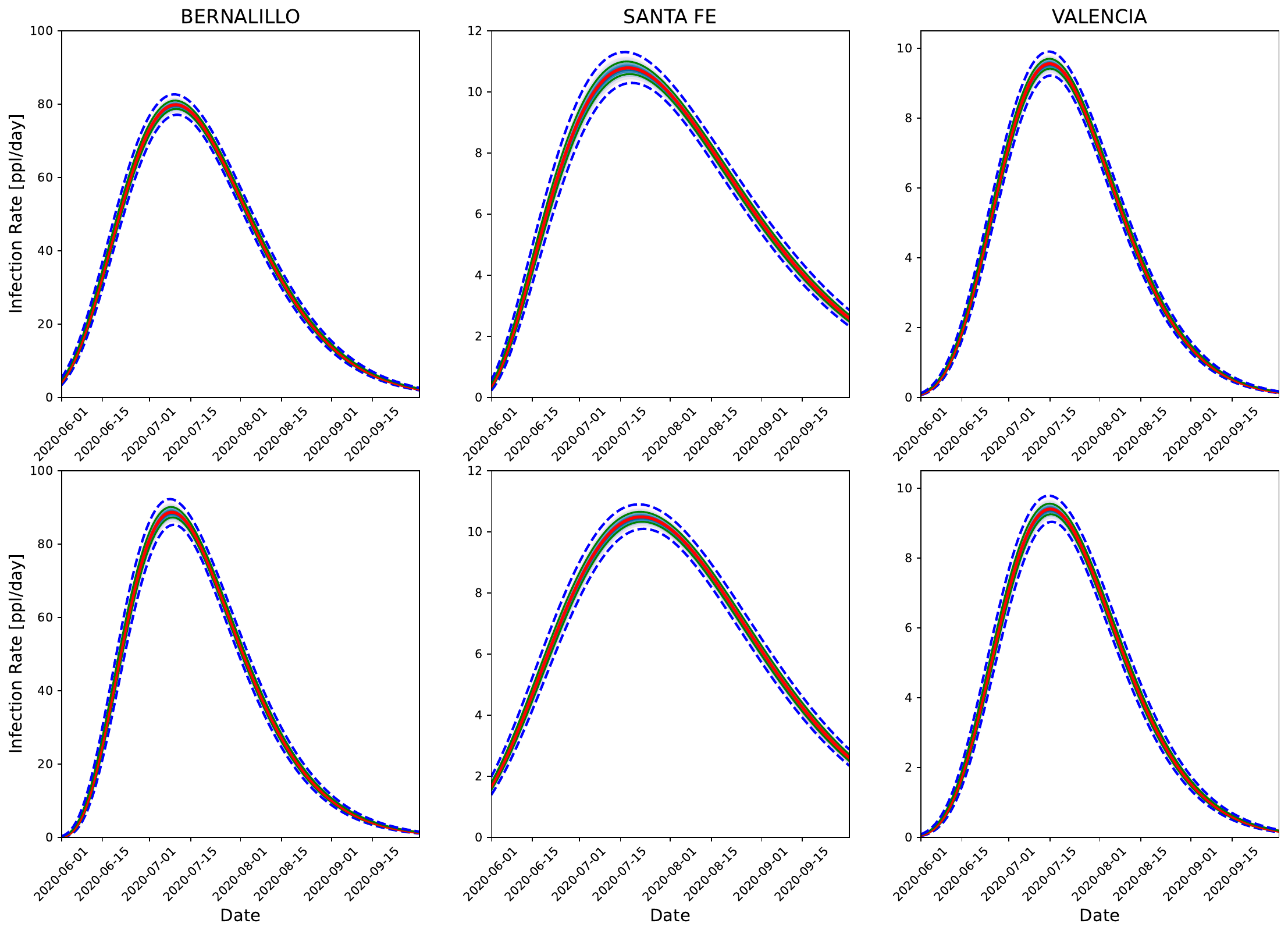}
	\caption{Comparison of the infection rate curves that determine predictions in Fig.~\ref{fig:3_county_pp}. These are for a 3-county inference of Bernalillo (left), Santa Fe (middle), and Valencia (right) done jointly using the GMRF model (top) and independently for each county (bottom). The red line shows the median predictions, the shaded green line shows the inter-quartile range, and the dashed lines are the $5^{\text{th}}$ and $95^{\text{th}}$ percentiles. Not much of a difference can be discerned visually.}
	\label{fig:3_county_fGamma_jo_in}
\end{figure}

\S~\ref{sec:3_county} describes the estimation of the infection-rate field over Bernalillo, Santa Fe and Valencia using MFVI jointly (where the GMRF spatial model is used) and individually. Fig.~\ref{fig:3_county_pp} plots the PPT runs, driven by a distribution of infection-rate fields. The corresponding fields are in Fig.~\ref{fig:3_county_fGamma_jo_in}.

\begin{figure}[!b]
	\centering
    \includegraphics[width=\textwidth]{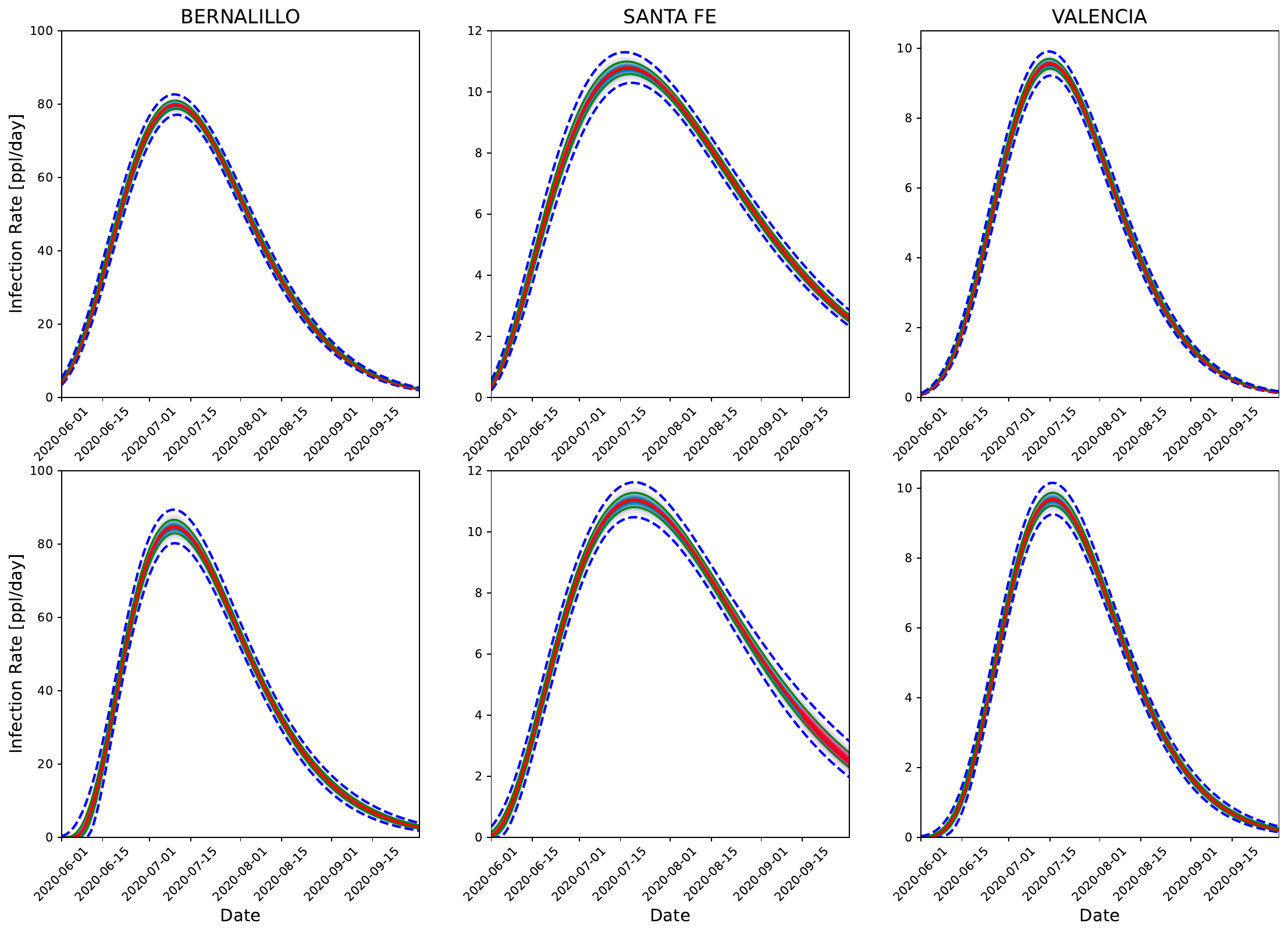}
	\caption{Comparison of the infection rate curves that determine predictions in Fig.~\ref{fig:3_county_pp_vimcmc}. These are for a 3-county inference of Bernalillo (left), Santa Fe (middle), and Valencia (right) done jointly using the MFVI (top) and AMCMC (bottom), taken from our Part I paper\cite{24sr3a,safta2024detecting} The red line shows the median predictions, the shaded green line shows the inter-quartile range, and the dashed lines are the $5^{\text{th}}$ and $95^{\text{th}}$ percentiles. Not much of a difference can be discerned visually.}
	\label{fig:3_county_fGamma_vi_mcmc}
\end{figure}

The same section describes the estimation of the infection-rate field over Bernalillo, Santa Fe and Valencia using AMCMC and the MFVI. Fig.~\ref{fig:3_county_pp_vimcmc} plots the PPT runs, driven by a distribution of infection-rate fields. The corresponding fields are in Fig.~\ref{fig:3_county_fGamma_vi_mcmc}.

\end{document}